\documentclass[aps,pra,twocolumn,amsmath,superscriptaddress,longbibliography]{revtex4-2}
\usepackage{graphicx}
\usepackage{adjustbox}
\usepackage{bm}
\usepackage{color}
\usepackage{braket}
\usepackage{multirow}
\usepackage{tikz}
\usepackage{mathrsfs}
\usepackage{dsfont}
\usepackage{comment}
\usepackage{amsmath}
\usepackage{mathtools}
\usepackage{booktabs}
\usepackage{caption}
\usepackage{subcaption}
\usepackage{float}  
\usepackage{amssymb}
\usepackage[urlcolor=blue,colorlinks=true,citecolor=blue,linkcolor=blue,pdfstartview={FitH},bookmarks=false]{hyperref}
\newcommand{\nn}{\nonumber}

\newcommand{\kuo}[1]{\left( #1 \right)}
\newcommand{\Sn}[1]{\left| \mathcal{S}_{#1} \right>}

\newcommand{\suppmatfootnote}{\footnote{See Supplemental Materials for more details, including (I) derivation of exact nonthermal eigenstates in XXZ magnets,  (II) entanglement of scar states, (III) expansion of SHS under the basis of scar states, (IV) local SU(2) algebra in XXZ magnets, (V) generalization of SU(2) algebra and scar dynamics to higher dimensions, (VI) the characteristics of initial states in the non-scar subspace under the krylov method.}}

\begin{document}

\title{Krylov complexity in quantum many-body scars of spin-1 models}

\author{Qingmin Hu}
\affiliation{College of Physics, Nanjing University of Aeronautics and Astronautics, Nanjing, 211106, China}
\affiliation{Key Laboratory of Aerospace Information Materials and Physics (NUAA), MIIT, Nanjing 211106, China}

\author{Wen-Yi Zhang}
\affiliation{College of Physics, Nanjing University of Aeronautics and Astronautics, Nanjing, 211106, China}
\affiliation{Key Laboratory of Aerospace Information Materials and Physics (NUAA), MIIT, Nanjing 211106, China}

\author{Yunguang Han}
\affiliation{College of Computer Science and Technology, Nanjing University of Aeronautics
and Astronautics, Nanjing 211106, China}
 
\author{Wen-Long You}
\email{wlyou@nuaa.edu.cn}
\affiliation{College of Physics, Nanjing University of Aeronautics and Astronautics, Nanjing, 211106, China}
\affiliation{Key Laboratory of Aerospace Information Materials and Physics (NUAA), MIIT, Nanjing 211106, China}

\begin{abstract}
Weak ergodicity breaking, particularly through quantum many-body scars (QMBS), has become a significant focus in many-body physics. Krylov state complexity quantifies the spread of quantum states within the Krylov basis and serves as a powerful diagnostic for analyzing nonergodic dynamics. In this work, we study spin-one XXZ magnets and reveal nonergodic behavior tied to QMBS. For the XY model, the nematic N\'eel state exhibits periodic revivals in Krylov complexity. In the generic XXZ model, we identify spin helix states as weakly ergodicity-breaking states, characterized by low entanglement and nonthermal dynamics. Across different scenarios, the Lanczos coefficients for scarred states display an elliptical pattern, reflecting a hidden SU(2) algebra that enables analytical results for Krylov complexity and fidelity. These findings, which exemplify the rare capability to characterize QMBS analytically, are feasible with current experimental techniques and offer deep insights into the nonergodic dynamics of interacting quantum systems.
\end{abstract}

\maketitle

\section{INTRODUCTION}
Ergodicity, a fundamental concept in statistical mechanics,  is still not fully understood in quantum mechanics. In classical systems, ergodicity implies that a system will explore its entire phase space over time and states eventually become uniformly distributed. However, its applicability in the quantum realm remains a subject of ongoing research. The eigenstate thermalization hypothesis (ETH)~\cite{ETHRigol2008,ETHPhysRevE.50.888,ETHPhysRevA.43.2046} posits that for generic closed quantum many-body systems, the long-time behavior of observables can be described by the thermal ensemble and indicates that the system reaches thermal equilibrium through the internal interactions.

However, there are important exceptions, such as integrable systems, where an extensive number of conserved quantities constrains the dynamics and prevents thermalization. Beyond integrable systems, chaotic systems can also challenge ergodicity through special mechanisms. For instance, many-body localization~\cite{MBLnp,MBLaop,MBLprb,MBLprl}  occurs in certain disordered systems, where interactions fail to induce thermalization due to the localization of excitations. Similarly, quantum many-body scars \cite{scarPhysRevLett.131.190401,scarPhysRevA.110.043312,scarAdler2024,scarPhysRevB.100.184312,scarPhysRevB.101.024306,scarPhysRevB.102.085140,scarPhysRevLett.123.030601,scarPhysRevB.102.224303,scarPhysRevB.107.L201105,scarPhysRevLett.125.230602,scarPhysRevLett.127.150601,scarPhysRevLett.128.090606,longlived,scarMoudgalya_2022,scarPhysRevB.110.134441, Serbyn2021, Zhang2023,PRR.4.013103,PRB.108.104411}
represent another unique phenomenon where nonthermal scar eigenstates coexist within a thermal spectrum, leading to persistent dynamics that violate ergodicity. The emergence of QMBS in a system depends strongly on the initial state, making it a key example of weak ergodicity breaking.  A few years ago, experiments on Rydberg atoms revealed unconventional quench dynamics when initiated from an antiferromagnetic state~\cite{Bernien2017}. This phenomenon, interpreted as QMBS within the PXP model~\cite{Turner2018,Turner2018_n}, has since been recognized as a form of weak ergodicity breaking. QMBS refers to nonthermal eigenstates that stand apart from the vast majority of thermal states in a quantum system. Extensive research on QMBS has identified several defining characteristics, including the existence of infinitely long-lived quasiparticle states, periodic revivals of many-body observables, and low-entanglement stationary states exhibiting uniform level spacing~\cite{doi:10.1146/annurev-conmatphys-031620-101617}. 

Krylov state complexity \cite{PhysRevD.106.046007,SpreadCaputa2023,SpreadPhysRevB.109.014312,SpreadDixit2024,SpreadCamargo:2024deu,SpreadNizami:2024ltk}, a powerful tool for studying quantum dynamics, has recently gained attention for its ability to provide insights into phenomena such as chaos, integrability, and weak ergodicity breaking in quantum systems. Originally introduced in the context of quantum circuits, quantum complexity is closely related to Nielsen complexity~\cite{Nielsen,NielsenE.105.064117,Nielsenspin}, which measures the minimum number of quantum gates needed to transform one quantum state into another. Krylov complexity, developed further in the framework of quantum adiabatic dynamics, comes in two forms: the operator version, which quantifies the growth of a local operator during adiabatic evolution through nested commutators with the Hamiltonian~\cite{Kcop}. The state version, also known as spread complexity, tracks how the components of an initial quantum state evolve and spread over a dynamically generated Krylov subspace~\cite{PhysRevD.106.046007, Time_e}. This framework has been especially useful in understanding the dynamics of QMBS and identifying the initial states responsible for nonthermal behavior within an otherwise thermal system. Recent studies, such as those on the PXP model~\cite{PRB_sc,Nandy_2024}, have shown how Krylov state complexity can be used to examine weak ergodicity breaking, with the closure of the Krylov subspace and distinct patterns in Lanczos coefficients revealing the role of local SU(2) symmetry in shaping QMBS dynamics.

In this work, we explore weak ergodicity breaking in spin-1 quantum systems through the  {character} of Krylov complexity. By focusing on the spin-1 XY and XXZ magnets, we uncover a novel interplay between local SU(2) algebra constraints and quantum many-body scars. Specifically, our study reveals how nematic N\'eel states in XY magnets exhibit periodic revivals in Krylov complexity, and identifies the spin helix state (SHS) in XXZ magnets as exhibiting distinct non-ergodic dynamics. These findings not only extend the known mechanisms of quantum many-body scars but also analytically reveal Krylov complexity as a powerful diagnostic tool for understanding non-ergodic behavior in interacting systems.

The paper is organized as follows: In Sec. \ref{kc}, we briefly review Krylov complexity for states and introduce the Lanczos algorithm. In Sec. \ref{XY}, we analyze the behavior of Lanczos coefficients and Krylov complexity in spin-1 XY magnets. Section \ref{XXZ} focuses on the SHS, a specific class of eigenstates in the spin-1 XXZ model. We describe the QMBS in spin-1 XXZ magnets and examine the corresponding behavior of Lanczos coefficients and Krylov complexity from the SHS. Finally, Sec. \ref{con} provides a summary of our findings.

\section{KRYLOV COMPLEXITY}\label{kc}
We begin with a concise description of the Lanczos algorithm and its relation to Krylov complexity, which quantifies the spread of quantum states over time. Consider a unitary evolution under a time-independent Hamiltonian:
\begin{eqnarray}
    \left| \psi \left( t \right) \right>  & =e^{-i\hat{H}t}\left| \psi \kuo{0}\right> ,
\end{eqnarray}
where $\hat{H}$ is a time-independent Hamiltonian and $ \left| \psi(0) \right> $ is the initial state. For very small $t$, the state $ \left| \psi(t) \right> $  can be approximated as $ \left| \psi(0) \right> - i \hat{H} t\left| \psi(0) \right> $. As time progresses, higher powers of the Hamiltonian $\hat{H}$ are involved, resulting in a linear superposition of states from the set  $\left\{ \left| \psi(0) \right> , \hat{H} \left| \psi(0) \right> , \hat{H}^2\left| \psi(0) \right>, \cdots \right\} $.  However, this set does not form an orthogonal basis, as orthogonality is not guaranteed.
Thus, the Gram-Schmidt procedure is applied to the set $\left\{ \hat{H}^n\left| \psi \kuo{0}\right> \right\} $ and generates an orthonormal basis known as the Krylov basis, $\left\{ \left| \mathcal{K}_n \right> \right\}$:
\begin{eqnarray}
 &&\left| \mathcal{K}_0 \right> =\left| \psi \kuo{0} \right>,  b_0=0, \nonumber \\ 
 && {\hat{H}\ket{\mathcal{K}_{n}} = b_{n}\left| \mathcal{K}_{n-1} \right> + a_{n}\ket{\mathcal{K}_{n}} + b_{n+1}\left| \mathcal{K}_{n+1} \right>,} 
\end{eqnarray}
where $a_n$=$\left< \mathcal{K}_n \right|\hat{H}\left| \mathcal{K}_n \right>$,  {$b_n=\left< \mathcal{K}_n \right|\hat{H}\left| \mathcal{K}_{n-1} \right>$} and $\left< \mathcal{K}_n \mid \mathcal{K}_n \right>$ $= 1$.
This process will generate the Lanczos coefficients  $a_n$, $b_n$ sequently. The procedure is typically truncated when $b_K \rightarrow 0 $, at which point the dimension of the Krylov subspace is reached.
In this Krylov subspace, the Hamiltonian is represented by a tridiagonal matrix:
\begin{eqnarray}
    \hat{H}_{\mathcal{K}}=
    \left( 
        \begin{matrix}
        a_0&		b_1&		0&		\cdots&		0\\
        b_1&		a_1&		b_2&		\cdots&		0\\
        0&		b_2&		a_2&		\cdots&		0\\
        \vdots&		\vdots&		\vdots&		\ddots&		b_{K-1}\\
        0&		0&		0&		b_{K-1}&		a_{K-1}\\
        \end{matrix} 
    \right).
\end{eqnarray}
The evolved state can be expressed in terms of this Krylov basis:
\begin{eqnarray}\label{krylovbasis}
    \left| \psi \left( t \right) \right> =\sum_{n=0}^{K-1}{\phi_n\left( t \right)}\left| \mathcal{K}_n \right>, 
\end{eqnarray}
where unitarity requires $\sum_{n=0}^{K-1}{\left| \phi _n\left( t \right) \right|^2}=1$. 
By applying the Schr\"{o}dinger equation  to this expression and using the recursion relations from the Lanczos coefficients $a_n$ and $b_n$,  we obtain the following time evolution:
\begin{eqnarray}\label{tri-diagona}
    i\partial _t\phi _n\left( t \right) =a_n\phi _n\left( t \right) +b_{n+1}\phi _{n+1}\left( t \right) +b_n\phi _{n-1}\left( t \right) .
\end{eqnarray}
This reveals that any Hamiltonian can be mapped to a one-dimensional chain,
where each site $n$ has a local potential $a_n$ and interacts only with its nearest neighbors through hopping terms $b_n$ and $b_{n+1}$.
Most importantly, in the Krylov basis,  {the operator
that quantifies Krylov complexity is defined as follows}~\cite{PhysRevD.106.046007}: 
\begin{eqnarray}
    \hat{C}_{\mathcal{K}} = \sum_{n=0}^{K-1} n \left|\mathcal{K}_n\right\rangle \left\langle\mathcal{K}_n\right|.
\end{eqnarray}
Thus, Krylov complexity is given by 
\begin{eqnarray}\label{C_k}
    C_{\mathcal{K}}\left( t \right) = \left< \psi \kuo{t}\right| \hat{C}_{\mathcal{K}} \left| \psi\kuo{t} \right>= \sum_{n=0}^{K-1}{n\left| \phi _n\left( t \right) \right|^2}.
\end{eqnarray}
 From Eq. (\ref{C_k}), Krylov complexity represents the average position of the wavefunction in the Krylov subspace. This can be understood analogously to the Schr\"{o}dinger equation for a single particle hopping along a half-infinite chain, where the basis states correspond to positions of the chain and the coefficients $\left| \phi_n\left( t \right) \right|^2$ represent the probability distribution over these positions.  {At $t=0$, the particle starts from $\phi_0$, and as time evolves, it hops away from $n=0$,}  consequently reducing the fidelity and increasing the Krylov complexity. Inspired by this connection, it is realized that the dynamics is governed by the Lanczos coefficients $\{b_n\}$. Hence, if there exist certain universal behaviors of fidelity, it is conceivable that their information is hidden inside $\{b_n\}$.
 
 We study a generic spin-1 XXZ chain described by the Hamiltonian
\begin{eqnarray}
\label{Hamhxxz}
\hat{H}_{\rm{XXZ}} = J \sum_{ {\langle i,j \rangle}} \hat{I}_{i,j}(\Delta) + h \sum_j \hat{S}_j^z, 
\end{eqnarray}
where $\hat{I}_{i,j}(\Delta) = \hat{S}_i^x \hat{S}_j^x + \hat{S}_i^y \hat{S}_j^y + \Delta \hat{S}_i^z \hat{S}_j^z$ describes the XXZ interactions with anisotropy $\Delta$ along the $z$-axis between sites $i$ and $j$. In Eq.(\ref{Hamhxxz}), $\hat{S}_j^\mu$ ($\mu= x, y, z$) are spin-1 operators at site $i$ on a $d$-dimensional hypercubic lattice of volume $\mathcal{V} = L^d$. The parameter $h$ 
represents the strength of the magnetic field along the $z$-axis.  {The sum $\sum_{\langle i,j \rangle}$ runs over nearest-neighbor pairs of lattice sites.}

The eigenstates of the spin-1 operators $\hat{S}_j^\mu$ are denoted by $\vert m_j^\mu \rangle$, where $m_j^\mu = 1, 0, -1$ are the corresponding eigenvalues of  $\hat{S}_j^\mu$ along the  { $\mu$-axis}. It is convenient to work with a “$p$-wave basis” of local Hilbert space, i.e., 
\begin{eqnarray}
\vert x_j\rangle\!&=&\!\frac{1}{\sqrt{2}}(\vert m_j^z=-1\rangle\!-\!\vert m_j^z=1\rangle), \nonumber \\
\vert y_j\rangle\!&=&\!\frac{1}{\sqrt{2}}(\vert m_j^z= -1\rangle\!+\!\vert m_j^z=1\rangle),\nonumber \\
\vert z_j\rangle\!&=&\!\vert m_j^z=0\rangle.
\end{eqnarray}  In this representation, we have 
\begin{eqnarray}
\label{Smatrices}
      S_j^x&=&
     \left( 
        \begin{matrix}
            0& 0& 0\\
            0& 0& 1\\
            0& 1& 0
        \end{matrix} 
    \right),
    S_j^y=
    \left( 
        \begin{matrix}
            0& 0& -i\\
            0& 0& 0\\
            i& 0& 0
        \end{matrix} 
    \right),  S_j^z=
    \left( 
        \begin{matrix}
        0& 1& 0\\
            1& 0& 0\\
            0& 0& 0
        \end{matrix} 
    \right).\quad \quad
\end{eqnarray}
It is obvious that $ \left|x_j\right>$ is the eigenstate of $\hat{S}_j^x$ with eigenvalue 0, i.e., $ \left|x_j\right> = \left|m_j^x = 0\right>$, and $ \left|y_j\right>$ is the eigenstate of $\hat{S}_j^y$ with eigenvalue 0, i.e., $ \left|y_j\right> = \left|m_j^y = 0\right>$.

\section{SCARS IN SPIN-1 XY MAGNETS} \label{XY}
We begin by considering a $d$-dimensional spin-1 XY model with $\Delta=0$. The Hamiltonian is given by:
\begin{eqnarray} 
\label{HamXY} 
\hat{H}_{\rm XY}&=& J\sum_{\left< ij \right>} (\hat{S}_i^{x}\hat{S}_j^{x}+\hat{S}_i^{y}\hat{S}_j^{y}) + h\sum_j \hat{S}_j^{z},
\end{eqnarray}
where $\left< ij \right>$ denotes nearest-neighbor pairs. While the spin-1/2 XY chain can be solved exactly using the Jordan-Wigner transformation, the spin-1 XY model with external magnetic field  is a well-known non-integrable system. 

In Eq. \eqref{HamXY}, several low-entanglement eigenstates, known as scar states, have been reported to exhibit weak ergodicity breaking~\cite{PhysRevLett.123.147201}. These scar states exist for both open boundary conditions (OBCs) and periodic boundary conditions (PBCs), for any values of $h$ and even system size $L$. We define the generator of the SU(2) algebra as:
\begin{eqnarray} \label{Jpm} \hat{J}^{\pm} = \frac{1}{2} \sum_j e^{i \boldsymbol{r}_j \cdot \boldsymbol{\pi}} \left( \hat{S}_j^{\pm} \right)^2, \end{eqnarray}
where $\boldsymbol{r}_j$ represents the coordinates of the spins, and $\boldsymbol{\pi} = (\pi, \pi, \dots, \pi)$ is a vector with all components equal to $\pi$. The scar states are given by:
\begin{eqnarray} 
\label{nbimagnons}\left| \mathcal{S}_l \right> = \mathcal{N}_l \left( \hat{J}^+ \right)^l \left| \Omega \right>, \end{eqnarray}
where $l = 0, 1, \dots, \mathcal{V}$ is an integer, and $\left| \Omega \right> = \bigotimes_j \left| m_j^z = -1 \right>$ is a fully polarized state. The normalization factors are  $\mathcal{N}_l = \sqrt{(\mathcal{V}-l)!/l! \mathcal{V}!}$.  
The corresponding energies of these scar states are $E_l = h(2l -\mathcal{V}) $, which do not depend on the coupling strength $J$ since the exchange term annihilates all scar states~\cite{PhysRevLett.123.147201}.

\begin{figure}[tb]
\centering
\includegraphics[width=1.0\columnwidth]{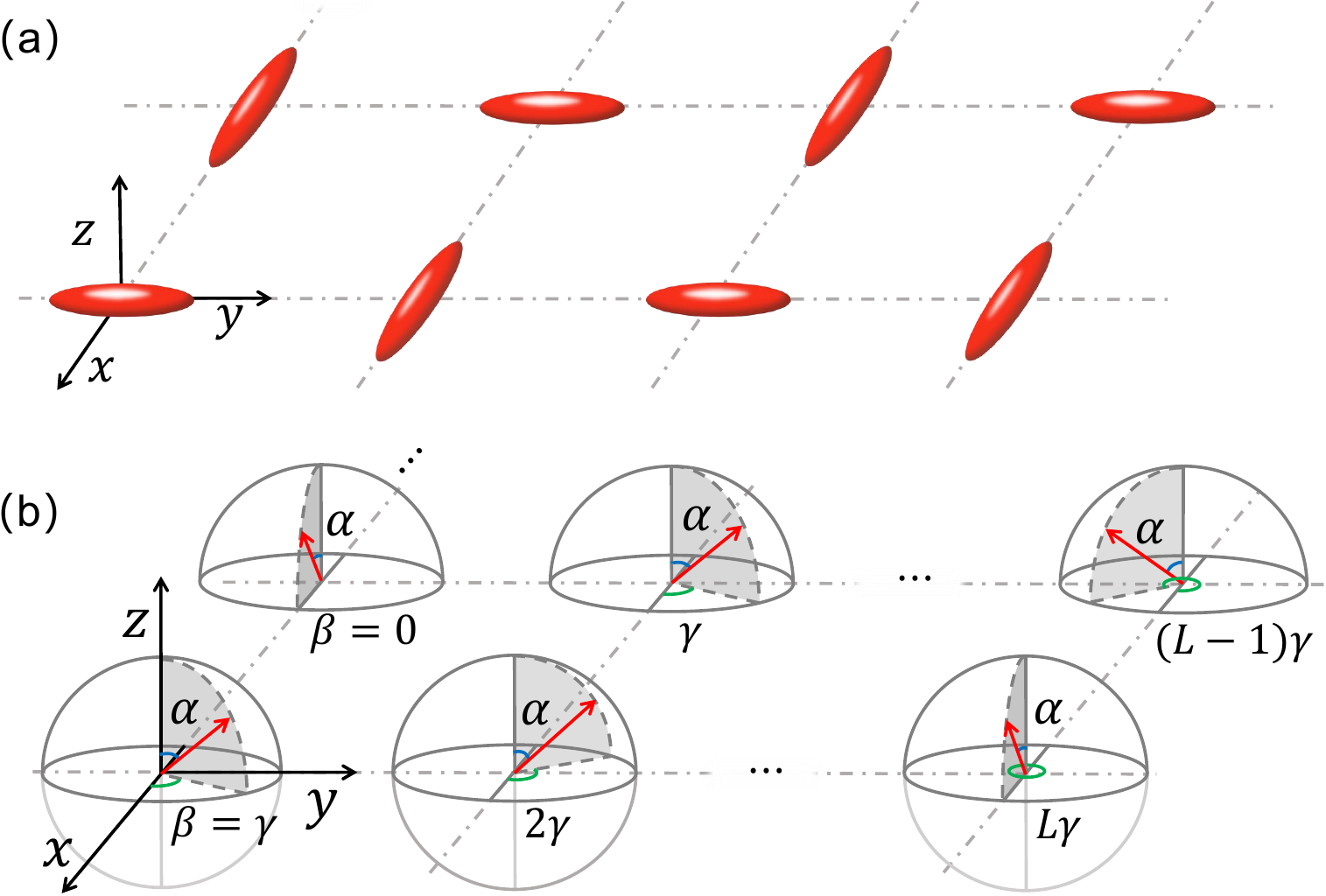} 
    \caption{\raggedright Schematic of the nematic N\'eel state and spin helix state on a $d=2$ dimensional square lattice. (a) The spin structure of the nematic N\'eel state, where the red ellipse represents the spin state $\left|m_j^\mu=0\right>$ ($\mu=x$ or $y$), and the long orientation indicates its $a$-axis. (b) The spin structure of the helix state, where the red arrows represent the spin state $\left|\alpha,\beta\right>$ in Eq.(\ref{coherentstate_alpha_beta}). Here, $\alpha$ is the polar angle measured from the $z$-axis, and $\beta = j\gamma$ is the azimuthal angle measured from the $x$-axis in the $xy$-plane, with lattice index $j = 1, \cdots , L$.}
    \label{neelhelix}
\end{figure}

Then, a superposition of  the scar  states in Eq.(\ref{nbimagnons}), which contains $l$ bimagnons  each with momentum $ \boldsymbol{\pi}$,  can be used to construct the  {nematic} N\'eel state $\left| \mathrm{NN} \right>$, 
written as 
 {
\begin{eqnarray}
\left| \mathrm{NN} \right> 
=\sum_{l=0}^{\mathcal{V}}{\sqrt{\frac{\mathbb{C}_{\mathcal{V}}^l}{ 2^\mathcal{V}}} \left| \mathcal{S}_l \right>},
\end{eqnarray}
where the notation $\mathbb{C}_{\mathcal{V}}^l$ denotes the binomial coefficient. }
It can be verified that\begin{eqnarray} \label{neelstate}
 \left| \mathrm{NN} \right> & =& \bigotimes_{j=1}^{\mathcal{V}}{\left( \frac{\left| m_j^z=+1 \right> - e^{i\boldsymbol{r}_j \cdot \boldsymbol{\pi}}\left| m_j^z=-1 \right>}{\sqrt{2}} \right)}.
\end{eqnarray}
It can be further simplified into $ \vert   y x y x  \cdots \rangle$, which corresponds to a set of spin-nematic directors precessing in the $x$-$y$ plane, as shown in Fig.  \ref{neelhelix}(a).

\emph{ SU(2) algebra in $XY$ models.--}
We now examine the behavior of the Lanczos coefficients in XY magnets. 
Without loss of generality, we focus on one-dimensional lattice ($d=1$). The Hamiltonian can be decomposed as follows:
\begin{eqnarray} 
\label{HJ1}
\hat{H}_J&=& J\sum_{j} \left(\hat{S}_j^{x}\hat{S}_{j+1}^{x}+\hat{S}_j^{y}\hat{S}_{j+1}^{y}\right), \\
\label{Hh1}
\hat{H}_{h}&=&   h\sum_j \hat{S}_j^{z}.
\end{eqnarray}
For any $l$, we have $ \hat{H}_J \left| \mathcal{S}_l \right> = 0$~\cite{PhysRevLett.123.147201}.
Consequently, since the nematic N\'eel state $\left| \mathrm{NN} \right>$ can be expressed as a linear combination of $\left| \mathcal{S}_l \right>$ states, it is also an eigenstate of Eq.(\ref{HamXY}). This implies
\begin{eqnarray} \label{HJHD}
    \hat{H}_J (\hat{H}_{\rm XY})^{n} \left| \mathrm{NN} \right> = 0, \quad n\ge 0,
\end{eqnarray}
indicating that $\hat{H}_J$ acts as zero in the subspace spanned by the Krylov basis vectors built from  $\left| \mathrm{NN} \right>$.
 
The remaining term $\hat{H}_{h}$ represents the contribution from the Zeeman field.
Since  $S_j^z$ in Eq.(\ref{Smatrices}) interchanges the  $\vert x_j\rangle$ and $\vert y_j\rangle$ states, 
 we can map spin-1 states to effective spin-1/2 states for simplicity. Under this mapping, we represent the nematic N\'eel state $\left| \mathrm{NN} \right>$ as a product state of alternating spins, $\left| \downarrow \uparrow \cdots \right\rangle$,  where $\left| \downarrow \right>$  represents  $\left| y \right>$ and $\left| \uparrow \right>$ corresponds to   $\left| y \right>$  on odd and even sites, respectively. To facilitate later usage, we apply a $\pi$-rotation about the $x$-axis on the even sites, which swaps the states  $\left| \uparrow \right>$ and $\left| \downarrow \right>$, transforming the nematic  N\'eel  state into a ferromagnetic configuration. 
  
 {In this context,} the Zeeman field term in Eq.~(\ref{Hh1}) can be reformulated as an effective spin-1/2 Hamiltonian:
\begin{eqnarray}
\hat{H}_{\rm eff,\frac{1}{2}} = h \left(\hat{\mathfrak{h}}^{+}  + \hat{\mathfrak{h}}^{-}\right). 
\end{eqnarray}
where \begin{eqnarray} 
\hat{\mathfrak{h}}^{+} &=& \sum_{j=1}^{L/2}{\sigma_j^{+}}= \sum_{j=1}^{L/2} \left( \left|x_{2j-1}\right> \left<y_{2j-1}\right| + \left|y_{2j}\right> \left<x_{2j}\right| \right) \nonumber
 \end{eqnarray}
 is the raising operator, with the corresponding lowering operator $\hat{\mathfrak{h}}^{-}$ defined analogously. These operators, $\hat{\mathfrak{h}}^{+}$ and $\hat{\mathfrak{h}}^{-}$ , form a pair of generators for the SU(2) algebra, satisfying the commutation relations:
\begin{eqnarray} 
[\hat{\mathfrak{h}}^{z}, \hat{\mathfrak{h}}^{\pm}] = \pm \hat{\mathfrak{h}}^{\pm}, \quad [\hat{\mathfrak{h}}^+, \hat{\mathfrak{h}}^-] = 2\hat{\mathfrak{h}}^z,
 \end{eqnarray}
 where the projection operator $\hat{\mathfrak{h}}^{z}$ is given by 
\begin{eqnarray} 
\hat{\mathfrak{h}}^{z}  
&=& \frac{1}{2}\sum_{j=1}^{L/2}\left( \left|x_{2j-1}\right>\left<x_{2j-1}\right| - \left|y_{2j-1}\right>\left<y_{2j-1}\right| \right. \nonumber \\ &+& \left. \left|y_{2j}\right>\left<y_{2j}\right| - \left|x_{2j}\right>\left<x_{2j}\right| \right). 
\end{eqnarray}

In this SU(2) representation,  we can adopt spin states $\left| S, m \right>$, which are eigenstates of the total spin operator $\vec{\hat{\mathfrak{h}}}^2$ defined as
\begin{eqnarray}
\vec{\hat{\mathfrak{h}}}^2=\frac{1}{2} \left(\hat{\mathfrak{h}}^{+} \hat{\mathfrak{h}}^{-} + \hat{\mathfrak{h}}^{-} \hat{\mathfrak{h}}^{+}\right) + (\hat{\mathfrak{h}}^z)^2, 
\end{eqnarray}
where $S$  represents the total spin and $m$ is the spin projection along the $z$-axis. 
It is straightforward to verify that
\begin{eqnarray}
&& \hat{\mathfrak{h}}^z \left| \mathrm{NN} \right> = -\frac{L}{2} \left| \mathrm{NN} \right>, \vec{\hat{\mathfrak{h}}}^2 \left| \mathrm{NN} \right> = \frac{L}{2} \left( \frac{L}{2} + 1 \right) \left| \mathrm{NN} \right>. \quad \quad
\end{eqnarray}
Thus, the nematic N\'eel state $\left| \mathrm{NN} \right>$
corresponds to the state $\left| S, -S \right>$, equivalent to $\left| \downarrow \downarrow \cdots \downarrow \right>$ with $S = L/2$. 
The action of the effective Hamiltonian $\hat{H}_{\rm eff, \frac{1}{2}}$ on the state $\left|S,-S+n\right>$ is:
\begin{eqnarray}
 &&\hat{H}_{\rm eff,\frac{1}{2}} \left|S,-S+n\right>  \nonumber \\
 &=&h \sqrt{(n+1)\left(L-n\right)} \left| S,-S+n+1\right>\nonumber \\
&+&h \sqrt{n\left(L+1-n\right)} \left|S,-S+n-1\right>.  
\end{eqnarray}
By orthogonalizing the sequence
$\{ \left|S, -S\right>, $ 
$\hat{H}_{\rm eff , \frac{1}{2}} \left|S, -S\right>, $ 
$\kuo{\hat{H}_{\rm eff , \frac{1}{2}}}^2 \left|S,-S\right>,$ 
$\cdots \} $, we obtain the Krylov basis $\left\{ \mathcal{K}_n \right\} $, where
\begin{eqnarray}
\label{Krylovbasis}
    \left\{  \left| \mathcal{K}_n \right> \right\}=\left\{  \left| S,-S+n \right>  \right\} , n=0,1,2,\cdots,2S.
\end{eqnarray}
The Lanczos coefficients are:
\begin{align}
    a_n &= \left< \mathcal{K}_n \right| \hat{H}_{\rm eff,\frac{1}{2}} \left| \mathcal{K}_n \right> = 0, \\
    b_n &= \left< \mathcal{K}_{n} \right| \hat{H}_{\rm eff,\frac{1}{2}} \left| \mathcal{K}_{n-1} \right> = h\sqrt{n\left(2S+1-n\right)}.   \label{eqbn}
\end{align}

\begin{figure}[t!]
\includegraphics[width=\columnwidth]{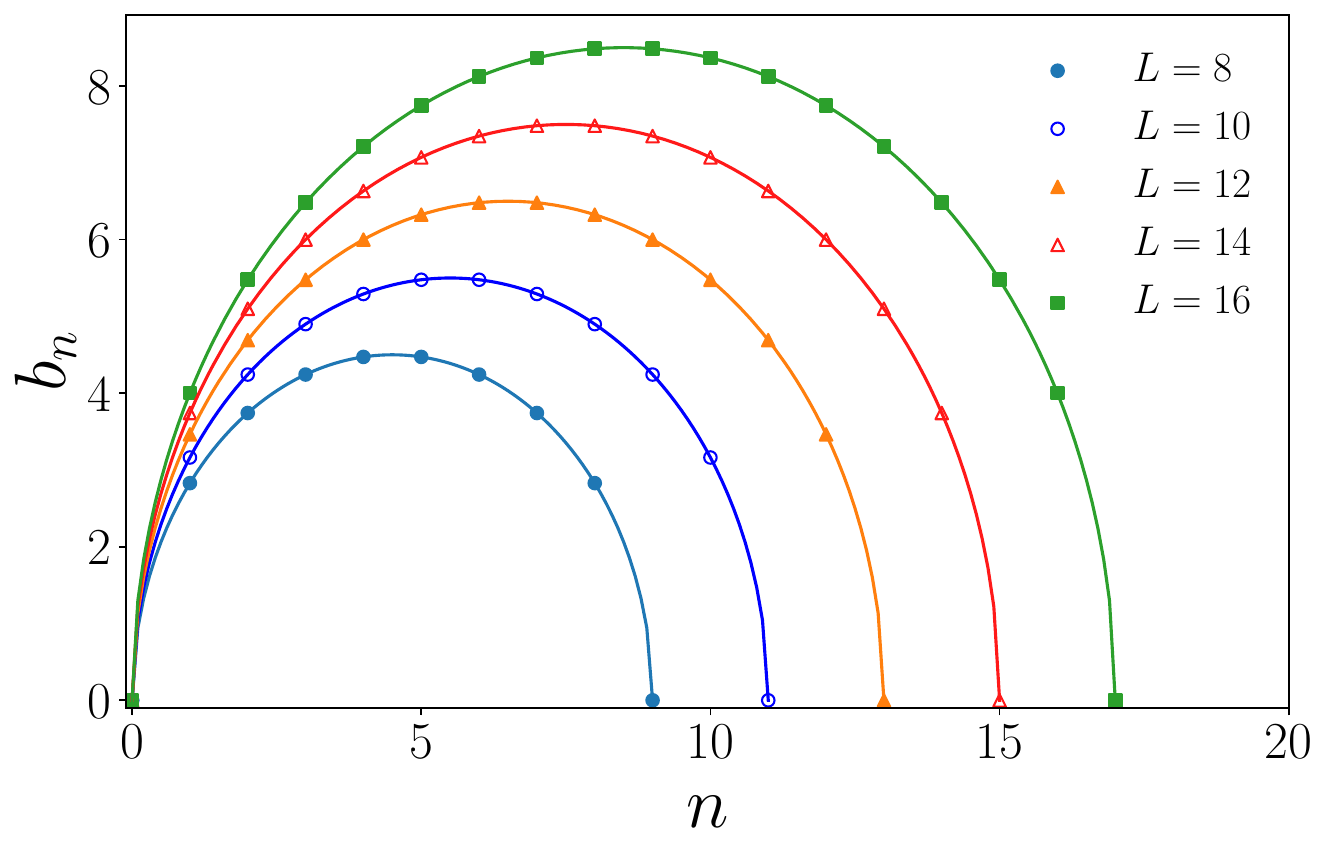}
\caption{\raggedright The Lanczos coefficients  $b_n$ versus Krylov index $n$ for one-dimensional nematic N\'eel state \eqref{neelstate}. The dots represent the numerical results, while the solid line corresponds to the analytical expression in Eq.~\eqref{eqbn}. The agreement between the numerical data and the analytical expression is excellent for all system sizes $L$.}
\label{bn}
\end{figure}
In Fig.~\ref{bn}, we plot the Lanczos coefficients $b_n$ as a function of Krylov index $n$ for the nematic N\'eel state.  The function $b_n$ versus $n$ forms an elliptical curve  with a minor axis of length $L+1$.  Notably, $b_{n}$ reaches zero at $n=L+1$, indicating that the Krylov subspace  closes prematurely.
This suggests that the evolution of the initial state is fully captured by the first $L+1$ Krylov basis vectors, consistent with the fact that Eq. (\ref{neelstate})  spans exactly $L+1$ eigenstates.

We then analyze the quench dynamics from the initial state in the spin-$S$ representation. In the scarred subspace, Eq.~(\ref{HJHD}) shows that $ \hat{H}_J $ contributes a constant zero term, representing the non-dynamical contribution in this regime. The time evolution of the state is given by:
\begin{eqnarray}
\left| \psi \left(t\right) \right> =  e^{-i\hat{H}_{\rm eff,\frac{1}{2}}  {t} }\left| S,-S \right>,
\end{eqnarray} 
where the time evolution operator, using the Baker–Campbell–Hausdorff (BCH) formula, is expressed as:
\begin{eqnarray}
    e^{-i\hat{H}_{\rm eff,\frac{1}{2}}t}=
    e^{c_+ \hat{\mathfrak{h}}^{+}}e^{c_0 \hat{\mathfrak{h}}^{z}}e^{c_- \hat{\mathfrak{h}}^{-}}\;
\end{eqnarray}
 {with the coefficients $c_+(t)=c_-(t)= -i \tan (ht )$ and 
$c_0(t)=-2\log\left[\cos (ht)\right]$}.
The time-evolved state then becomes
\begin{eqnarray}
&& \left|{\psi(t)}\right>=
e^{-Sc_0}\sum^{2S}_{n=0}c_+^n
\sqrt{\mathbb{C}_{2S}^n}
\left|{S,-S+n}\right>.  \label{psit}
\end{eqnarray}

\begin{figure}[t!]
    \includegraphics[width=\columnwidth]{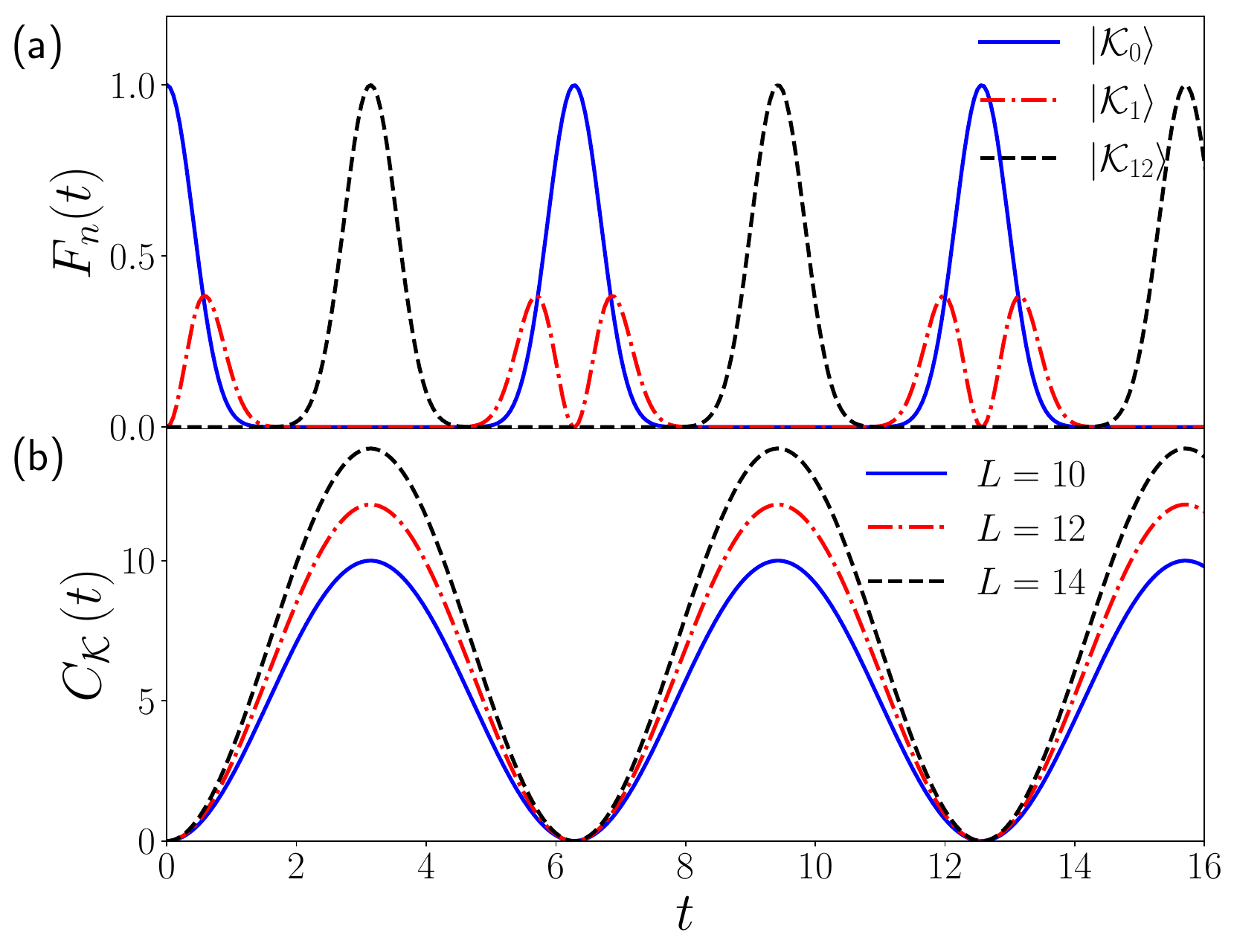}    \caption{\raggedright The dynamics of   one-dimensional XY model in (\ref{HamXY}) starting from nematic N\'eel state (\ref{neelstate}).  The system size  $L = 12$, and the parameters are $J = 1$ and $h = 0.5$. (a) The fidelity between the Krylov vectors $\{ \left| \mathcal{K}_n \right>$, $n$ = 0, 1, $L$ $\}$ and the time-evolved state. (b) The Krylov complexity of the nematic N\'eel state for various system sizes.}
    \label{FCxy}
\end{figure}

Starting from the initial state $\left| \mathrm{NN} \right>$, the fidelity shows periodic revivals. This behavior indicates that the system retains information from the initial state over extended time scales and demonstrates a clear violation of the ETH. The fidelity between the time-evolved state and the Krylov basis in Eq.~(\ref{psit}) is given by:
\begin{eqnarray}
F_n(t) = \left| \left\langle \psi(t) \mid \mathcal{K}_n \right\rangle \right|^2 =  \cos^{2L}(ht)     \tan^{2n}(ht)  
\mathbb{C}_L^n.  
\end{eqnarray}
In Fig.~\ref{FCxy}(a), we show the probability of finding individual Krylov basis vectors during the evolution. Starting from the initial state $\left| \mathcal{K}_0 \right>$, the state that exhibits the same oscillation pattern at later times is $\left| \mathcal{K}_L \right>$,
which  corresponds to  $\left|S,S\right>$ in this Krylov basis.  This is the fully polarized state: 
\begin{eqnarray}
\label{KL}
        \left| \mathcal{K}_L \right> &= & \left|S,S\right>=\left| \uparrow \uparrow \cdots \right>.
        \end{eqnarray}
This state, which can also be written as $ \vert  x y x y  \cdots \rangle$, is effectively a one-site translation of $\left| \mathcal{K}_0 \right>$.

We continue our analysis of weak ergodicity breaking in spin-1 XY magnets through Krylov complexity, focusing on the Lanczos coefficients.  The coefficients of the time-evolved state (\ref{psit}) in the Krylov basis  are given by:
\begin{eqnarray}
    \phi_n(t)=e^{-Sc_0}c_+^n\sqrt{\mathbb{C}_{2S}^n}.
\end{eqnarray}
Using this form, the Krylov complexity, defined in Eq.~(\ref{C_k}), takes the form: 
\begin{eqnarray}
C_{\mathcal{K}}(t)=2S\sin^2\left(ht\right),
\end{eqnarray}
which exhibits periodic oscillations in time. As shown in Fig.~\ref{FCxy}(b), the Krylov complexity oscillates at the same frequency as the state fidelity. The complexity reaches its maximum value $L$, when the time-evolved state $\left| \psi(t) \right> $ coincides with the Krylov basis vector $\left| \mathcal{K}_{L} \right> $.

\section{SCARS IN SPIN-1 XXZ MAGNETS} \label{XXZ}
We now direct our focus to the spin-1 XXZ model,  {which} provides a more generalized framework by incorporating anisotropy along the $z$-axis. The corresponding Hamiltonian is given in Eq. \eqref{Hamhxxz}.  
While the spin-1/2  XXZ model is exactly solvable via the Bethe ansatz \cite{Bethzbook_1982,Bethzbook_1999}, the spin-1 XXZ model is generally not integrable. However, in specific limits, simplifications arise. At the isotropic point ($\Delta=1$), the model reduces to the spin-1 Heisenberg chain, which, although not exactly solvable, exhibits the Haldane gap. In the Ising-like limit ($\Delta \to \infty$), the model can be approximated  through perturbative or semiclassical methods. For intermediate  values of anisotropy, numerical techniques such as density matrix renormalization group (DMRG) or exact diagonalization are typically employed to study the associated many-body dynamics. 
The spin-1/2 XXZ model has revealed many significant phenomena in ultracold atoms, including quantum many-body scars and spin transport \cite{Jepsen2020}. Notably, a long-lived phantom helix state has been observed \cite{longlived,Zhang_2024,1_2helix}. 

Motivated by these findings, we propose a new SHS in the spin-1 model. Through a combination of analytical and numerical techniques, we demonstrate that this state exhibits a relatively long relaxation time, which we attribute to the presence of scar eigenstates.  
We have identified a new class of quantum many-body scars distinct from those described in the Sec. \ref{XY}. These scars emerge when the anisotropy parameter satisfies $\Delta = \cos{\gamma}$, where $\gamma = 2q\pi/L$ with integer values $q=0,1,2,\cdots,L-1$. These specific values of $\gamma$ correspond to quantized modes within the system. The eigenstates associated with these quantum many-body scars are given by:
\begin{eqnarray}\label{Scarl}
    \left| \mathcal{S}_l \right> = \mathcal{L}_l  \left( \hat{S}_{\gamma}^+ \right) ^l \left| \Omega \right>,
\end{eqnarray}
where $\mathcal{L}_l = \sqrt{ (2\mathcal{V}-l)!/l!(2\mathcal{V})!}$ is a normalization constant, and  $\left| \Omega \right> = \bigotimes_j \left| m_j^z = -1 \right>$ is a fully polarized state. The  operator
 $ {\hat{S}_{\gamma}^+} = \sum_j{e^{-i \boldsymbol{r}_j \cdot  \boldsymbol{\gamma} }\hat{S}_j^+}$, 
 where $\boldsymbol{\gamma} = (\gamma, \gamma, \dots, \gamma)$ is a vector with all components equal to $\gamma$. 
 Each action of the operator $\hat{S}_{\gamma}^+$ excites a magnon, increasing the total magnetization $M^z$  by 1. Specifically, for the scar eigenstate $\Sn{l}$, the total magnetization is given by $M^z = l-\mathcal{V} $. In addition to affecting magnetization,  {$\hat{S}_{\gamma}^+$} also generates a translationally invariant state. 
 Since the initial state $\Sn{0}$ has zero momentum, the momentum of the scar state $\Sn{l}$ becomes $\boldsymbol{k}=-l\boldsymbol{\gamma}$.

\begin{figure}[t!]
\centering
\includegraphics[width=\columnwidth]{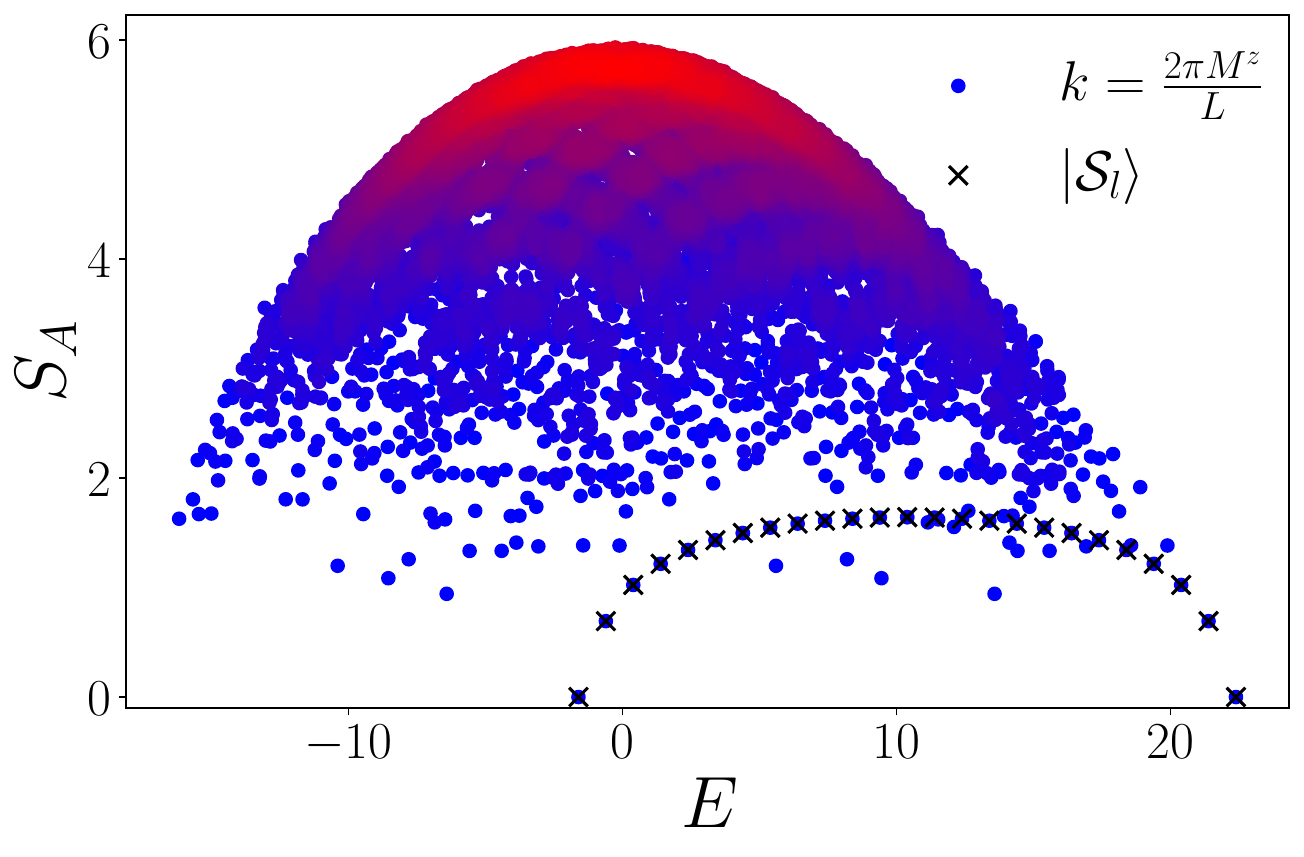}
   \caption{\raggedright
Bipartite entanglement entropy $S_A$ versus eigenenergies $E$ in a one-dimensional XXZ magnet. The system size is $L = 12$, and the model parameters are $J = 1$, $h = 1$, and $\Delta = \cos(2\pi /L)$. The energy spectrum is plotted in the scar subspaces of momentum $k = 2\pi M^z/L$. The scar eigenstates $\left| \mathcal{S}_l \right>$ are marked by black crosses.}
\label{spectrum xxz}
\end{figure}

In Fig.~\ref{spectrum xxz}, the bipartite entanglement entropy $S_A$ is plotted against the eigenenergies $E$ for the one-dimensional XXZ model. The vertical axis represents the bipartite entanglement entropy, defined as $S_A = -\mathrm{Tr}_\mathrm{B} \rho_A \ln \rho_A$, where the lattice is partitioned into two equal halves, and $\rho_A$ is the reduced density matrix of one half.  The energy spectrum is organized into scarred subspaces labeled by momentum  $k={2q\pi M^z}/{L}$, which arise from translational invariance and respecting both U(1) symmetry and translational symmetry.  Within these subspaces, the eigenstates marked by black crosses represent the scarred eigenstates $\Sn{l}$, which are distinguished by their low entanglement entropy and nearly uniform energy spacing. Specifically, the two polarized states, $\Sn{0}$ and $\Sn{2L}$, have zero entanglement entropy, while the entanglement entropy of the intermediate states $\Sn{l}$ increases sequentially, peaking at $\Sn{L}$. These scarred eigenstates form a unique set of nonthermal states that coexist with high-entanglement thermal eigenstates, providing evidence of weak ergodicity breaking in the system.  The eigenvalue equation for these scarred states in the one-dimensional case is given by \suppmatfootnote: 
 \begin{eqnarray} \label{eigeneq} \hat{H}_{\rm XXZ}\Sn{l} = \left[ J\Delta L + (l-L)h \right] \Sn{l}.
 \end{eqnarray}



We then show that the SHS, composed of the scar eigenstates in Eq. \eqref{Scarl}, plays a central role in the dynamics of quantum scars.  The SHS is a product state of local spin coherent states, each twisted by a phase factor along the $z$-axis, which is defined as:
\begin{eqnarray}
\label{generalSHS}
    \left|\mathrm{SHS}\kuo{\alpha,\gamma}\right> = \bigotimes^{\mathcal{V}}_{j=1}{ \hat{\mathcal{D}}_z\left(\boldsymbol{r}_j \cdot \boldsymbol{\gamma}\right)\left|\alpha,0\right>_{j}},
\end{eqnarray}
where $\hat{ \mathcal{D}}_z \left(\beta \right) = \exp\left(-\beta \hat{S}_j^{z} \right)$ is the rotation operator by $\beta$ along the $z$-axis.  
The spin coherent state corresponding to the operator $\hat{S}_j^n=\hat{\boldsymbol{S}}_j\cdot \boldsymbol{n}$ is given by:
\begin{eqnarray}  
\label{coherentstate_alpha_beta}\left| \alpha,\beta \right> &=& \cos^2 \left( \frac{\alpha}{2} \right) e^{-i\beta }\left|m^z=1\right> + \frac{1}{\sqrt{2}}\sin\alpha \left|m^z=0\right> \nonumber \\ 
        &+& \sin^2 \left( \frac{\alpha}{2} \right) e^{ i\beta }\left|m^z=-1\right>.
\end{eqnarray}
Here $\boldsymbol{n} = \left( \sin\alpha\cos\beta, \sin\alpha\sin\beta, \cos\alpha \right)$, where $\alpha$ is the polar angle measured from the $z$-axis, and $\beta$ is the azimuthal angle measured from the $x$-axis within the $xy$-plane. The coefficients in the coherent state (\ref{coherentstate_alpha_beta}) are normalized to ensure a consistent description of the spin state.  
For $\alpha=0$, the vector points along the $z$-axis. When  $\alpha>0$, $\ket{\mathrm{SHS}\kuo{\alpha,\gamma}}$ represents a spin helix configuration where the phase twist $\gamma$ evolves across the lattice sites, as illustrated in Fig. \ref{neelhelix}(b).

In the following, we focus on the specific case of a one-dimensional lattice ($d=1$), without loss of generality. In this scenario, the SHS simplifies to: 
\begin{eqnarray} \label{helix} \left|\mathrm{SHS}(\alpha,\gamma)\right> 
= \bigotimes^{L}_{j=1}{\left|\alpha,j\gamma\right>_j}.
\end{eqnarray} 
The SHSs are eigenstates of the XXZ model (\ref{Hamhxxz}) in the absence of an external magnetic field. The corresponding eigenvalue equation in this case is
\begin{eqnarray}
\hat{H}_{\Delta}\left|\mathrm{SHS}\kuo{\alpha,\gamma}\right> = J\Delta L   \left|\mathrm{SHS}\kuo{\alpha,\gamma}\right>, \end{eqnarray}
where 
$\alpha$ an arbitrary parameter, and  $\gamma =\arccos\Delta$. Here, the Hamiltonian is given by
\begin{eqnarray}
&&\hat{H}_{\Delta} = J\sum_{j} {\left( \hat{S}_j^x \hat{S}_{j+1}^x + \hat{S}_j^y \hat{S}_{j+1}^y + \Delta \hat{S}_j^z \hat{S}_{j+1}^z \right)}.  \end{eqnarray}
While the SHS in Eq. \eqref{helix} is not an eigenstate of the XXZ model for arbitrary values of $\Delta$, it is important to note that the SHS $\ket{\mathrm{SHS}\kuo{\alpha,\gamma}}$  can be expressed entirely as a superposition of the scar eigenstates $\ket{\mathcal{S}_l}$ from Eq. \eqref{Scarl}. This relationship is explicitly given by \footnotemark[\value{footnote}]:
\begin{eqnarray} \label{form}
\ket{\mathrm{SHS}\left(\alpha,\gamma\right)} = e^{iq\kuo{L+1}\pi}\sum_{l=0}^{2L}{\frac{(\cos{\frac{\alpha}{2}})^{l}(\sin{\frac{\alpha}{2}})^{2L-l}}{l!\mathcal{L}_l}} \ket{\mathcal{S}_l}.  \quad
\end{eqnarray}

\begin{figure}[t!]
\includegraphics[width=1.0\columnwidth]{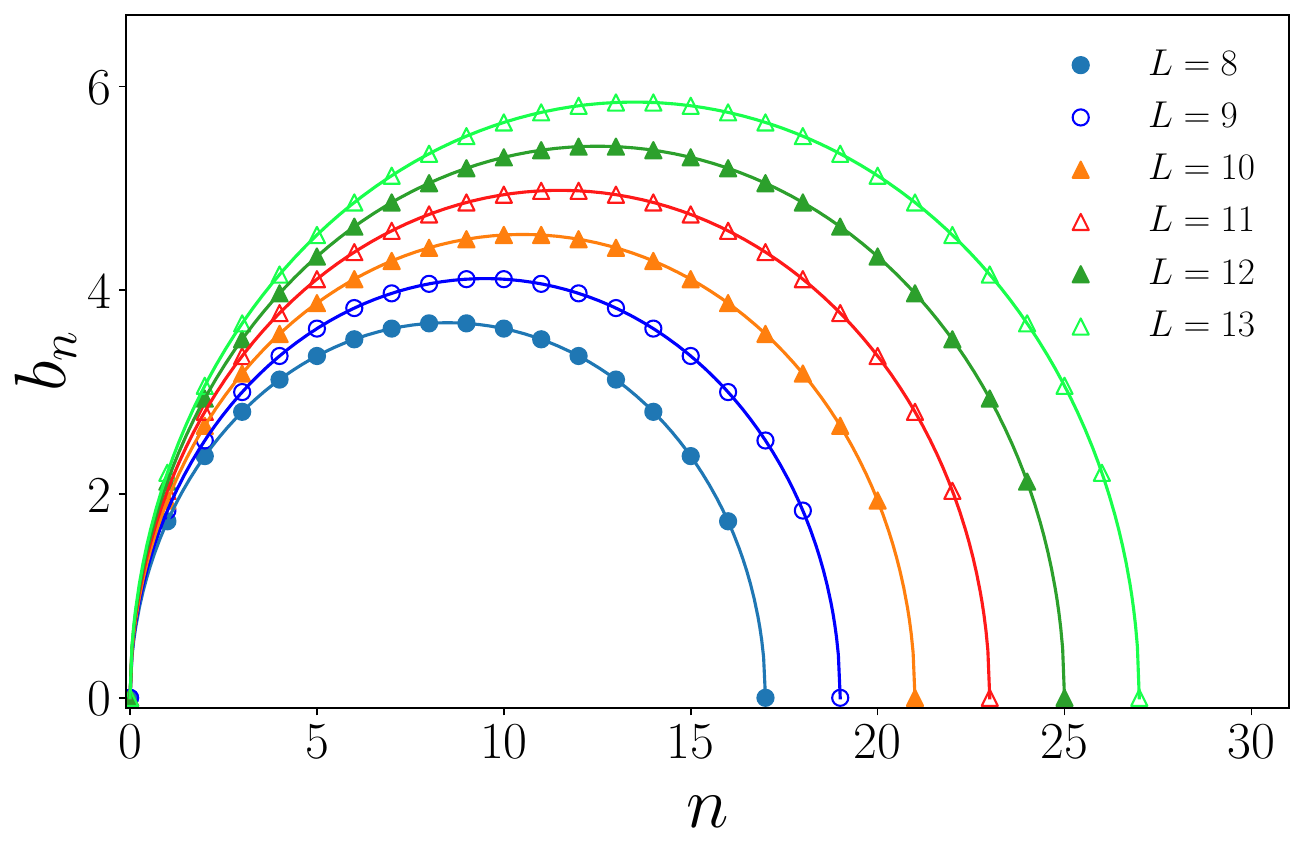}
\caption{\raggedright
Lanczos coefficients $b_n$  versus Krylov index $n$ of spin helix states $\ket{\mathrm{SHS}\kuo{\alpha,\gamma}} $ defined in Eq.~(\ref{generalSHS}) in one-dimensional XXZ magnets (\ref{Hamhxxz}) for various system sizes $L$. The model parameters are $J = 1$, $h = 1$, $\gamma = 2\pi /L$, and $\alpha = \pi/3$. The dots indicate the numerical results, while the line corresponds to Eq.~\eqref{bnxxz}. Both are in excellent agreement for all considered system sizes $L$.}
    \label{figbnxxz}
\end{figure}

When using $\ket{\mathrm{SHS}\kuo{\alpha,\gamma}}$ as the initial state, we observe the same behavior for the Lanczos coefficients as when initializing from the nematic N\'eel state $\left| \mathrm{NN} \right>$ in XY magnets.  The forms of $a_n$ and $b_n$ are derived in \footnotemark[\value{footnote}] , and the plot in Fig.~\ref{figbnxxz} confirms these results. The Lanczos coefficient $b_n$  exhibits an elliptical growth, given by:
\begin{eqnarray}\label{bnxxz}
    b_n = \frac{1}{2}h\sin{\alpha}\sqrt{n(2L+1-n)}.
\end{eqnarray}
In contrast, $a_n$ remains linear:
\begin{eqnarray}\label{anxxz}
    a_n = J\Delta L+ h\cos\alpha\left(L-n\right).
\end{eqnarray}
When $n=2L+1$, $b_n$ decays to zero, indicating that the Krylov subspace is closed, with a dimension of  $2L+1$.

The observed behavior reflects an underlying local SU(2) algebra, which allows us to derive exact solutions for key physical quantities. The fidelity between the time-evolved state and the Krylov basis is given by \footnotemark[\value{footnote}]: 
\begin{eqnarray}
\label{fidelity}
&& F_n(t) = \left| \left\langle \psi(t) \mid \mathcal{K}_n \right\rangle \right|^2 \nn \\
&&=\mathbb{C}_{2L}^n\kuo{1-\sin^2{\alpha}\sin^2\frac{ht}{2}}^{\kuo{2L-n}} \kuo{\sin^2{\alpha} \sin^2\frac{ht}{2}}^{n}.\quad\quad
\end{eqnarray}

\begin{figure}[t!]
\includegraphics[width=\columnwidth]{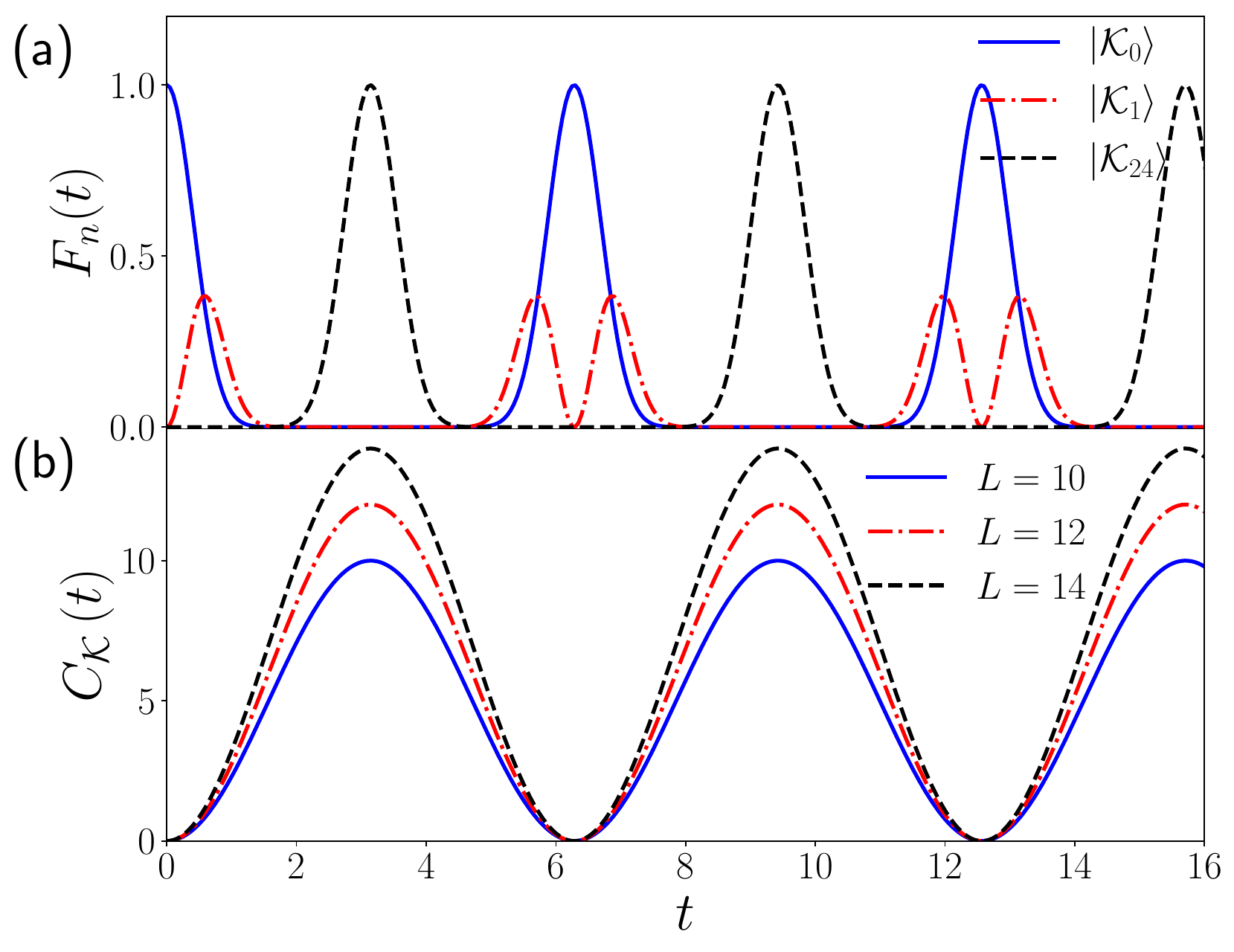}
\caption{\raggedright
The dynamics of the one-dimensional XXZ model (\ref{Hamhxxz}) starting from $\ket{\mathrm{SHS}\kuo{\alpha,\gamma}} $ defined in Eq.~(\ref{generalSHS}) with parameters $J = 1$, $h = 1$, $\alpha = \pi/2$, and $\gamma = 2\pi /L$. (a) The overlap of three Krylov vectors $\{$$ \left| \mathcal{K}_n \right>$, $n$=0, 1, $2L\}$ with the time-evolved state.  The system size $L = 12$. 
(b) The Krylov complexity of the spin helix state with repest to time $t$ for various system size $L$.}
    \label{FCxxz}
\end{figure}

The results are exhibited in Fig.~\ref{FCxxz}. The state exhibits periodic oscillations with a period of $ {\frac{2\pi}{h}}$, indicating a clear violation of thermalization. In Fig.~\ref{FCxxz} (a), in addition to the periodic revival of the initial state, a distinct oscillatory pattern is observed in the state $\ket{\mathcal{K}_{2L}}$. From this, we infer that $\ket{\mathcal{K}_{2L}} = \ket{\mathrm{SHS}\kuo{\alpha-\pi,\gamma}}$,  which indicates that $\ket{\mathcal{K}_{2L}}$ is also a SHS.

The Krylov complexity is expressed as \footnotemark[\value{footnote}] 
\begin{eqnarray}\label{complexity}
    C(t) = 2L \sin^2{\alpha} \sin^2{\frac{ht}{2}}. 
\end{eqnarray}
In Fig.~\ref{FCxxz}(b), the maximum Krylov complexity reaches $2L$, as the system fully occupies the state $\ket{\mathcal{K}_{2L}}$ at this point, contributing a weight of $2L$ to the complexity. Notably, the oscillation period is independent of system size.

In addition, under OBCs, the Lanczos coefficient $b_n$ remains unchanged, indicating that the transition strength is unaffected. However, the diagonal element $a_n$  experiences a positive shift by $J\Delta$. This shift does not result in imaginary energies, despite the presence of non-Hermitian boundary terms. As a result, the overall dynamical behavior remains consistent, and both  Eq.~\eqref{fidelity} and Eq.~\eqref{complexity}  remain applicable for OBCs, just as they are for PBCs .

\section{Discussion and CONCLUSION}\label{con}
\label{4}
In this work, we have systematically explored weak ergodicity breaking in spin-1 models using Krylov complexity. By analyzing the formalism of the Lanczos algorithm, we have provided a quantitative characterization of quantum many-body scars in both XY and XXZ magnets. We show that the hidden SU(2) symmetry in the XY and XXZ models enabled us to derive closed-form expressions for the Krylov complexity  and fidelity dynamics.

In the spin-1 XY model, the nematic  N\'eel state  [Eq.~\eqref{neelstate}]  was shown to exhibit periodic revivals of Krylov complexity. We identified that the Lanczos coefficients for this state display a characteristic elliptical pattern, which directly stems from an underlying SU(2) algebra. This symmetry constrains the dynamics to a closed Krylov subspace, with dimensionality determined by the number of scarred eigenstates. 
 Extending our analysis to the spin-1 XXZ model, we explored the system's behavior for a generic anisotropy parameter  $\Delta$ and identified a distinct class of scarred eigenstates at discrete values of $\Delta=\cos{\frac{2\pi q}{L}}$. In the thermodynamic limit ($L \to \infty$), these values become continuous. These scarred eigenstates form a basis for the spin helix state [Eq.~\eqref{generalSHS}], which we identified as a dynamically robust initial condition.
By deriving explicit expressions for the Lanczos coefficients and demonstrating their elliptic dependence on the Krylov index, we established a direct correspondence between the spectral properties of QMBS and Krylov dynamics. 

These results reveal a remarkable periodicity in the evolution of Krylov complexity, further confirming the nonergodic behavior of the system. The ability to analytically obtain these quantities shows the utility of Krylov complexity as a diagnostic tool for weak ergodicity breaking.  Notably, our work generalizes the concept of SHS from spin-1/2 systems~\cite{longlived} to spin-1 systems. These findings are particularly relevant for experimental platforms such as ultracold atoms~\cite{transport,Yang2020} and Rydberg systems~\cite{Rydbergscience.abg2530, RydbergKeesling2019}, where spin-1 dynamics can be realized and controlled. The robustness of the spin helix state's dynamics against thermalization, coupled with its tunability via the discrete anisotropy parameter $\Delta$, positions it as a promising candidate for exploring QMBS in programmable quantum simulators.   {
A notable discovery is the intriguing connection between Krylov complexity and the underlying SU(2) algebra, which provides novel strategies for designing quantum systems with weak ergodicity breaking. Our methods rely on Lie algebraic structures, suggesting a natural extension to systems with higher symmetries~\cite{PRR.2.043305}.
}

\begin{acknowledgments}
This work is supported by the National Natural Science Foundation of
China (NSFC) under Grant No. 12174194, Postgraduate Research \&
Practice Innovation Program of Jiangsu Province, under Grant No. KYCX23\_0347, Opening Fund of the Key Laboratory of Aerospace Information Materials and Physics (Nanjing University of Aeronautics and Astronautics), MIIT, Top-notch Academic Programs Project of Jiangsu Higher Education Institutions (TAPP), and stable supports for basic institute research under Grant No. 190101. 
\end{acknowledgments}

\bibliography{Ref}

\clearpage

\widetext
\setcounter{equation}{0}
\setcounter{figure}{0}
\setcounter{table}{0}
\setcounter{page}{1}
\setcounter{section}{0}
\setcounter{tocdepth}{0}
\renewcommand{\thefigure}{S\arabic{figure}}
\numberwithin{equation}{section} 
\begin{center}
{\bf \large Supplemental Material for “Krylov complexity in quantum many-body scars of spin-1 models” }
\end{center}

\section{Derivation of Exact Nonthermal Eigenstates in XXZ Magnets} \label{Sl}
In this section, we will demonstrate that $\Sn{l}$ is an eigenstate of Eq. \eqref{Hamhxxz}, with $\Delta=\cos(\gamma)$. It is straightforward to show that $\Sn{l}$ must be an eigenstate of the magnetic field term, $h\sum_{j=1}^L{\hat{S}_j^z}\Sn{l} = h(-L+l)\Sn{l}$. Our primary focus is on the impact of $\hat{H}_{\Delta}$ acting on $\Sn{l}$.
\subsection{$\Sn{1}$ satisfies the eigenvalue equation of XXZ magnets}
To demonstrate that the states $\Sn{l}$ are indeed eigenstates of $H$, we work in the local $S^z$ basis, where we label the local spin states,
\begin{eqnarray}
    \ket{m_j^z=\pm 1}=\ket{\pm}_j,\ket{m_j^z=0} = \ket{0}_j.
\end{eqnarray}
Then we can obtain that $\Sn{1}$ has the following form,
\begin{eqnarray}\label{S1}
    \Sn{1} = \mathcal{L}_1\sum_{j=1}^L{e^{i\varphi_j}  \sqrt{2}\left|\right. {\underbrace{- \ldots}_{j-1} 0 \underbrace{\ldots -}_{L-j}}}\left.  \right> ,
\end{eqnarray}
where $\varphi_j = -j\gamma$. 
We will assume PBCs. We now consider the action of $\hat{H}_J$ on the $\Sn{1}$.

\begin{eqnarray}
        \hat{H}_J \Sn{1} &&= J\mathcal{L}_1 \sum_{j=1}^L{\sqrt{2} e^{i\varphi_j}  \left(\left|\right. {\underbrace{- \ldots}_{j-2} 0 \underbrace{\ldots -}_{L-j+1}}\left.\right> + \left|\right. {\underbrace{- \ldots}_{j+1} 0 \underbrace{\ldots -}_{L-j-1}}\left.\right> \right)} \nn \\
        &&= J\mathcal{L}_1 \sum_{j=1}^L {\sqrt{2}\kuo{ e^{i\varphi_{j-1}} +e^{i\varphi_{j+1}}} \left|\right. {\underbrace{- \ldots}_{j-1} 0 \underbrace{\ldots -}_{L-j}}\left.\right>} \nn \\
        &&= 2 J\Delta\mathcal{L}_1  \sum_{j=1}^L{\sqrt{2} e^{i\varphi_{j}}  \left|\right. {\underbrace{- \ldots}_{j-1} 0 \underbrace{\ldots -}_{L-j}}\left.\right>} \nn \\
        &&= 2J\Delta \Sn{1},
\end{eqnarray}
where we use the relation $e^{i\varphi_{j-1}}+e^{i\varphi_{j+1}}= 2\cos{\gamma}e^{i\varphi_j}$ and $e^{i\varphi_{L}}=1$.

\begin{eqnarray}
    &\hat{H}_{\Delta} \Sn{1} = \hat{H}_J\Sn{1} + J\Delta \sum_{j=1}^{L} \hat{S}_j^z \hat{S}_{j+1}^z \Sn{1}  = J\Delta L\Sn{1}.
\end{eqnarray}
Evidently, $\Sn{1}$ is an eigenstate of Eq. \eqref{Hamhxxz},
\begin{eqnarray}
\hat{H}_{\rm{XXZ}}\Sn{1}=\left[J\Delta L + h\kuo{-L+1}\right]\Sn{1}.
\end{eqnarray}

\begin{table}[h]
\centering
\begin{subtable}[t]{0.3\textwidth}
    \centering
    \caption{}
    \label{tab:subtable1}
    \begin{tabular}{|c|c|c|}
        \hline
         & $u-1$ to $u+2$ & coefficient \\ \hline
        target & $\begin{matrix}  -&0&0&- \\ \end{matrix}$ & $4e^{i\varphi_{2u+1}}$ \\ \hline
        1 & $\begin{matrix}  0&-&0&- \\ \end{matrix}$ & $4Je^{i\varphi_{2u}}$ \\ \hline
        2 & $\begin{matrix}  -&0&-&0 \\ \end{matrix}$ & $4Je^{i\varphi_{2u+2}}$ \\ \hline
        3 & $\begin{matrix}  -&+&-&- \\ \end{matrix}$ & $2Je^{i\varphi_{2u}}$ \\ \hline
        4 & $\begin{matrix}  -&-&+&- \\ \end{matrix}$ & $2Je^{i\varphi_{2u+2}}$ \\ \hline
        5 & $\begin{matrix}  -&0&0&- \\ \end{matrix}$ & $4J(L-3)\Delta e^{i\varphi_{2u+1}}$ \\ \hline
        Sum &   & $4J\Delta L e^{i\varphi_{2u+1}}$ \\ \hline
    \end{tabular}
\end{subtable}
\hfill
\begin{subtable}[t]{0.3\textwidth}
    \centering
    \caption{}
    \label{tab:subtable2}
    \begin{tabular}{|c|c|c|}
        \hline
         & $u-1$ to $u+3$ & coefficient \\ \hline
        target & $\begin{matrix}  -&0&-&0&- \\ \end{matrix}$ & $e^{i\varphi_{2u+2}}$ \\ \hline
        1 & $\begin{matrix}  0&-&-&0&- \\ \end{matrix}$ & $4e^{i\varphi_{2u+1}}$ \\ \hline
        2 & $\begin{matrix}  -&-&0&0&- \\ \end{matrix}$ & $4Je^{i\varphi_{2u+3}}$ \\ \hline
        3 & $\begin{matrix}  -&0&0&-&- \\ \end{matrix}$ & $4Je^{i\varphi_{2u+1}}$ \\ \hline
        4 & $\begin{matrix}  -&0&-&-&0 \\ \end{matrix}$ & $4Je^{i\varphi_{2u+3}}$ \\ \hline
        5 & $\begin{matrix}  -&0&-&0&- \\ \end{matrix}$ & $4J(L-4)\Delta e^{i\varphi_{2u+2}}$ \\ \hline
        Sum &   & $4J\Delta L e^{i\varphi_{2u+2}}$ \\ \hline
    \end{tabular}
\end{subtable}
\hfill
\begin{subtable}[t]{0.3\textwidth}
    \centering
    \caption{}
    \label{tab:subtable3}
    \begin{tabular}{|c|c|c|}
        \hline
         & $u-1$ to $u+1$ & coefficient \\ \hline
        target & $\begin{matrix}  -&+&- \\ \end{matrix}$ & $2e^{i\varphi_{2u}}$ \\ \hline
        1 & $\begin{matrix}  0&0&- \\ \end{matrix}$ & $4Je^{i\varphi_{2u-1}}$ \\ \hline
        2 & $\begin{matrix}  -&0&0 \\ \end{matrix}$ & $4Je^{i\varphi_{2u+1}}$ \\ \hline
        3 & $\begin{matrix}  -&+&- \\ \end{matrix}$ & $2J(L-4)\Delta e^{i\varphi_{2u}}$ \\ \hline
        Sum &   & 2J$\Delta L e^{i\varphi_{2u}}$ \\ \hline
    \end{tabular}
\end{subtable}
\caption{\raggedright Action of $\hat{H}_{\Delta}$ on different configurations and resulting coefficients in $\Sn{2}$. (a) Target state with double adjacent $\ket{0}$ states. (b) Target state with double separate $\ket{0}$ states. (c) Target state with single $\ket{+}$ state.}
\label{table_Sl2}
\end{table}

\subsection{$\Sn{2}$ satisfies the eigenvalue equation}
The composition of $\Sn{2}$ has two configurations, a double excitation at a single site, and a single excitation at two sites. The former has a site in $\ket{+}$ and the latter has two sites in $\ket{0}$. As follow,
\begin{eqnarray}
        \Sn{2}&=\mathcal{L}_{2}\sum_{j<r}{4e^{i\kuo{\varphi_j+\varphi_r}} \left|\right.\underbrace{-\cdots}_{j-1}0\underbrace{-\cdots-}_{k-j-1}0\underbrace{\cdots-}_{L-k}}\left.\right> +\mathcal{L}_{2}\sum_{j}{2e^{i2\varphi_{j}} \left|\right. {\underbrace{- \ldots}_{j-1} + \underbrace{\ldots -}_{L-j}}\left.\right>},
\end{eqnarray}
 We will show the results of $\hat{H}_{\Delta}$ acting on different configurations in the following Table \ref{table_Sl2}, where we do not elaborate on $\mathcal{L}_l$. Row $0$ is the target state, and the 3rd column is its relative coefficient in $\Sn{2}$. Row $> 1$ list which configurations can be changed to the target state by $\hat{H}_{\Delta}$, and to the 3rd column is its relative coefficient after the action by $\hat{H}_{\Delta}$. 
We sum the coefficients of Row $> 1$ and find that no matter which configuration, the coefficient becomes J$\Delta L$ times the original after the action of $\hat{H}_{\Delta}$. Therefore, $\Sn{2}$ is the eigenstate of Eq. \eqref{Hamhxxz}, 
\begin{eqnarray}
       \hat{H}_{\rm{XXZ}}\Sn{2}=\left[J\Delta L + \kuo{-L+2}h\right]\Sn{2}.
\end{eqnarray}

\subsection{$l>2$, representation of $\Sn{l}$}
For $\Sn{l}$ with $l > 2$, we first explain the configurations in $\Sn{l}$. While maintaining in the magnetic quantum number, the double excitation at a single site and the single excitation at two distinct sites have different weights. Intuitively, for each additional $\ket{+}$ in the configuration, its weight is half that of the $\ket{0}\ket{0}$ configuration. This arises from a simple permutation problem.

We first need to understand the initial problem, which is how to obtain $\ket{+0---\cdots}$ from $\ket{\Omega}$ under the action of $\hat{S}_{\gamma}^+$. There are $3!/2$ possible paths: $\hat{S}_1^+ \hat{S}_1^+ \hat{S}_2^+ \ket{\Omega}$, $\hat{S}_1^+ \hat{S}_2^+ \hat{S}_1^+ \ket{\Omega}$, and $\hat{S}_2^+ \hat{S}_1^+ \hat{S}_1^+ \ket{\Omega}$. To obtain $\ket{000--\cdots}$, there are $3!$ possible paths.
\begin{eqnarray}
    \Sn{l}=\mathcal{L}_l\sum_{\boldsymbol{c}\kuo{l}} \left[ \frac{l!}{2^g} \prod_{j=1}^{g} \kuo{e^{-ic_j\gamma}\hat{S}_{c_j}^+}^2 \prod_{j=1}^{l-2g} \kuo{e^{-ic_j'\gamma}\hat{S}_{c_j'}^+} \right] \ket{\Omega},
\end{eqnarray}
where, $\boldsymbol{c}\kuo{l}$ is an array that records the types and positions of excitations in a configuration. $\boldsymbol{c}=\left( c_1,\cdots,c_g ,c_1',\cdots,c_{l-2g}' \right)$, where $c_i$ is the lattice coordinate of the double excitations, and $c_i'$ is the lattice coordinate of the single excitations. The parameter $g$ denotes the total number of double excitations.

\subsection{Strict proof of the eigenvalue equation for $\Sn{l}$}
In the previous sections, we provided an example-based proof of the eigenvalue equation \eqref{eigeneq} for $\Sn{l}$. Here, we will proceed with a strict proof.

\begin{eqnarray}
    \left[\hat{I}_{j-1,j}\kuo{\Delta}+\hat{I}_{j,j+1}\kuo{\Delta}, e^{-ij\gamma}\hat{S}_j^+ \right] = 
     e^{-ij\gamma}\kuo{-\hat{S}_{j-1}^+ \hat{S}_{j}^z +  \Delta \hat{S}_{j-1}^z \hat{S}_{j}^+ - \hat{S}_{j}^z \hat{S}_{j+1}^+ + \Delta \hat{S}_{j}^+  \hat{S}_{j+1}^z} ,
\end{eqnarray}
\begin{eqnarray} 
    \left[ \hat{H}_\Delta, \hat{S}_{\gamma}^+ \right] =&& \sum_{j=1}^L {\left[\hat{I}_{j-1,j}\kuo{\Delta} + \hat{I}_{j,j+1} \kuo{\Delta}, e^{-ij\gamma} \hat{S}_j^+ \right]} \label{d1} \nn \\ 
    =&& \sum_{j=1}^L \kuo{{-e^{-ij\gamma}\hat{S}_{j-1}^+ \hat{S}_{j}^z + \Delta e^{-i j\gamma} \hat{S}_{j-1}^z \hat{S}_{j}^+ - e^{-i\kuo{j-1}\gamma} \hat{S}_{j-1}^z \hat{S}_{j}^+ + \Delta e^{-i\kuo{j-1}\gamma} \hat{S}_{j-1}^+ \hat{S}_{j}^z } } \nn \\
    =&& i\sin{\gamma} \sum_{j=1}^L  e^{-ij\gamma} \hat{S}_{j}^+ \kuo{\hat{S}_{j+1}^z - \hat{S}_{j-1}^z } ,
\end{eqnarray}
\begin{eqnarray} \label{d2}
     \left[\left[ \hat{H}_\Delta, \hat{S}_{\gamma}^+ \right], \hat{S}_{\gamma}^+ \right] =&& i\sin{\gamma} \sum_{j=1}^L \left[ e^{-i\kuo{j-1}\gamma }\hat{S}_{j-1}^+ \hat{S}_{j}^z - e^{-i\kuo{j+1}\gamma} \hat{S}_{j}^z \hat{S}_{j+1}^+,e^{-ij\gamma}\hat{S}_j^+\right] 
     = 0 ,
\end{eqnarray}
\begin{eqnarray} \label{d3}
    \hat{H}_\Delta \kuo{\hat{S}_{\gamma}^+}^l \ket{\Omega} = \hat{S}_{\gamma}^+ \hat{H}_\Delta \kuo{\hat{S}_{\gamma}^+}^{l-1} \ket{\Omega} + \left[ \hat{H}_\Delta, \hat{S}_{\gamma}^+ \right] \kuo{\hat{S}_{\gamma}^+}^{l-1} \ket{\Omega} \nn \\
    =\hat{S}_{\gamma}^+ \hat{H}_\Delta \kuo{\hat{S}_{\gamma}^+}^{l-1} \ket{\Omega} +  \kuo{\hat{S}_{\gamma}^+}^{l-1} \left[ \hat{H}_\Delta, \hat{S}_{\gamma}^+ \right] \ket{\Omega}.
\end{eqnarray}

According to Eq.~\eqref{d1}, for any $j$, the action of $\hat{S}_{j}^z$ on the fully polarized state $\ket{\Omega}$ generates a negative sign, and the operator $\kuo{\hat{S}_{j+1}^z - \hat{S}_{j-1}^z}$ annihilates $\ket{\Omega}$:
\begin{eqnarray} \label{d1Omega}
\left[ \hat{H}_\Delta, \hat{S}_{\gamma}^+ \right] \ket{\Omega} = 0.
\end{eqnarray}
Substituting this into Eq.~\eqref{d3}, we obtain:
\begin{eqnarray}
\hat{H}_\Delta \kuo{\hat{S}_{\gamma}^+}^l \ket{\Omega} = \hat{S}_{\gamma}^+ \hat{H}_\Delta \kuo{\hat{S}_{\gamma}^+}^{l-1} \ket{\Omega} \nn \\ = \kuo{\hat{S}_{\gamma}^+}^l \hat{H}_\Delta \ket{\Omega} = J\Delta L \kuo{\hat{S}_{\gamma}^+}^l \ket{\Omega}.
\end{eqnarray}
Therefore, we have:
\begin{equation} \label{HDSl}
\hat{H}_\Delta \Sn{l} = J\Delta L \Sn{l}.
\end{equation}
Therefore, the eigenvalue equation \eqref{eigeneq} is proven. The proof process is not limited to spin-1 systems. It is anticipated that the $\Sn{l}$ we defined are eigenstates of the spin-$s$ XXZ magnet. However, the explicit form of the spin helix state for spin-$s$ and its connection to $\Sn{l}$ are not rigorously demonstrated in this work.

\section{ENTANGLEMENT OF SCAR STATES}
In this appendix, we explain the bipartite entanglement entropy of scar eigenstates $\Sn{l}$ and the scaling.
We define the form of generalized scar states, 
\begin{eqnarray}
    \ket{\mathbf{S}^L_l}=\frac{1}{ \sqrt{\mathbb{C}_{2L}^l} l!}\kuo{\sum_{j=1}^L{e^{-ij\gamma}\hat{S}_j^+}}^l \ket{\Omega}.
\end{eqnarray}
Arbitrary scar eigenstates for $l<L$, with $L$ is even, can be expressed as,
\begin{eqnarray}
    \Sn{l<L} =&& \frac{1}{ \sqrt{\mathbb{C}_{2L}^l} l!} \sum_{n=0}^l \left[ \mathbb{C}_l^n \kuo{ \sum_{j=1}^{L_1} e^{-ij\gamma} \hat{S}_j^{+}}^n \ket{\Omega}_A \otimes e^{-iL_1\gamma} \kuo{ \sum_{j=1}^{L_2} e^{-ij\gamma} \hat{S}_j^{+}}^{l-n} \ket{\Omega}_B \right] \nn \\
    =&& e^{-iL_1\gamma} \frac{1}{ \sqrt{\mathbb{C}_{2L}^l} l!} \sum_{n=0}^l \left[ \mathbb{C}_l^n \sqrt{\mathbb{C}_{2L_1}^n \mathbb{C}_{2L_2}^{l-n} } n! (l-n)!\ket{\mathbf{S}^{L_1}_n}_A \otimes \ket{\mathbf{S}^{L_2}_{l-n}}_B \right] \nn \\
    =&& e^{-iL_1\gamma}  \sum_{n=0}^l \left[\sqrt{\frac{ \mathbb{C}_{2L_1}^n \mathbb{C}_{2L_2}^{l-n} }{ \mathbb{C}_{2L}^l}  } \ket{ \mathbf{S}^{L_1}_n}_A \otimes \ket{\mathbf{S}^{L_2}_{l-n}}_B \right], \\
    \rho_A\kuo{l} =&& \mathrm{Tr}_B \Sn{l} \left<\mathcal{S}_l\right| = \sum_{n=0}^l\frac{\mathbb{C}_{2L_1}^n \mathbb{C}_{ 2L_2 }^{l-n} }{\mathbb{C}_{2L}^l } \ket{\mathbf{S}^{L_1}_n}_A \bra{\mathbf{S}^{L_1}_n}.
\end{eqnarray}
The eigenvalue spectrum of the reduced density matrix for $\ket{\mathcal{S}_{l \le 2L_1}}$, 
\begin{eqnarray}
    \lambda_n = \frac{\mathbb{C}_{2L_1}^n \mathbb{C}_{2L_2}^{l-n} }{\mathbb{C}_{2L}^l }, \qquad
    S_A\kuo{l} = -\sum_{n=0}^{l} \lambda_n \ln{\lambda_n}.
\end{eqnarray}
$\ket{\mathcal{S}_{L+n}}$ and $\ket{\mathcal{S}_{L-n}}$ possess particle-hole symmetry, hence their bipartite entanglement entropy corresponds to each other.

\section{expansion of SHS under the basis of scar states}
\label{Relationship}

Here, we demonstrate the form of the one-dimension spin helix state $\ket{\mathrm{SHS}\kuo{\alpha,\gamma}}$ with any angle $\alpha$ under the basis of the scar states, i.e., Eq.~\eqref{form}.

\begin{eqnarray}
    \ket{\mathrm{SHS}\kuo{\alpha,\gamma}} &=&\bigotimes_{j=1}^L{ \left[\cos^2 \left( \frac{\alpha}{2} \right) e^{-ij\gamma }\left|+\right>_j + \frac{1}{\sqrt{2}}\sin\alpha \left|0\right>_j + \sin^2 \left( \frac{\alpha}{2} \right) e^{ ij\gamma }\left|-\right>_j\right]} \nonumber \\ 
    &=& \sin^{2L}\kuo{\frac{\alpha}{2}} e^{i\frac{L\kuo{L+1}}{2}\gamma} \bigotimes_{j=1}^L{ \left[\frac{\cos^2 \left( \frac{\alpha}{2} \right)}{\sin^2 \left( \frac{\alpha}{2} \right)} e^{-i2j\gamma }\left|+\right>_j + \sqrt{2} 
         \frac{\cos\left( \frac{\alpha}{2} \right)}{\sin\left( \frac{\alpha}{2} \right)}  e^{-ij\gamma }\left|0\right>_j + \left|-\right>_j\right]} \nonumber \\ 
    &=& \sin^{2L}\kuo{\frac{\alpha}{2}} e^{iq\kuo{L+1}\pi} \bigotimes_{j=1}^L{\left[\frac{1}{2} \kuo{\frac{\cos \left( \frac{\alpha}{2} \right)}{\sin \left( \frac{\alpha}{2} \right) }e^{-ij\gamma} S_j^+ }^2 \ket{-}_j  + {\frac{\cos \left( \frac{\alpha}{2} \right)}{\sin \left( \frac{\alpha}{2} \right) }e^{-ij\gamma} S_j^+ } \ket{-}_j + \ket{-}_j\right]} \nonumber \\
    &=& \sin^{2L}\kuo{\frac{\alpha}{2}} e^{iq\kuo{L+1}\pi} \prod_{j=1}^L{\left[\frac{1}{2} \kuo{\frac{\cos \left( \frac{\alpha}{2} \right)}{\sin \left( \frac{\alpha}{2} \right) }e^{-ij\gamma} S_j^+ }^2  + {\frac{\cos \left( \frac{\alpha}{2} \right)}{\sin \left( \frac{\alpha}{2} \right) }e^{-ij\gamma} S_j^+ }+1\right]} \ket{\Omega} \nonumber \\
    &=& \sin^{2L}\kuo{\frac{\alpha}{2}} e^{iq\kuo{L+1}\pi}  \sum_{l=0}^{2L} \sum_{\boldsymbol{c}\kuo{l}} \left[\frac{\cos^l \left( \frac{\alpha}{2} \right)}{\sin^l \left( \frac{\alpha}{2} \right) } \frac{1}{2^g} \prod_{j=1}^{g} \kuo{e^{-ic_j\gamma}\hat{S}_{c_j}^+}^2\prod_{j=1}^{l-2g} \kuo{e^{-ic_j'\gamma}\hat{S}_{c_j'}^+} \right] \ket{\Omega} \nonumber \\
    &=& e^{iq\kuo{L+1}\pi}\sum_{l=0}^{2L}{\frac{(\cos{\frac{\theta}{2}})^{l}(\sin{\frac{\theta}{2}})^{2L-l}}{l!\mathcal{L}_l}} \ket{\mathcal{S}_l}.
\end{eqnarray}

\section{SU(2) algebra in XXZ magnets}
\label{appC}
We will explain the behavior of the Lanczos coefficient in XXZ magnets below. 
We first define the SU(2) generators as follows:
\begin{eqnarray}
    \hat{U}^+ &=& \sum_{j=1}^{L} {\left[ -\sin{\alpha} \hat{S}_j^z + \kuo{\cos{\alpha}\cos{(j\gamma)} - i\sin{(j\gamma)}} \hat{S}_j^x  + \kuo{\cos{\alpha}\sin{(j\gamma)} + i\cos{(j\gamma)}} \hat{S}_j^y \right]}, \\
\hat{U}^- &=& \left( \hat{U}^+ \right)^{\dag}, \\
\hat{U}_0 &=& \sum_{j=1}^{L} \left[\cos{\alpha} \hat{S}_j^z + \sin{\alpha} \cos{\left(j\gamma\right)} \hat{S}_j^x + \sin{\alpha} \sin{\left(j\gamma\right)} \hat{S}_j^y \right].
\end{eqnarray}
Now we have the following relation,
\begin{eqnarray}
    \left[\hat{U}_0 , \hat{U}^\pm\right]=\pm \hat{U}^\pm,\qquad \left[\hat{U}^+, \hat{U}^- \right]=2\hat{U}_{0}\;.
\end{eqnarray}
Next, we define the common eigenstates of $\hat{\boldsymbol{U}}^2$ and $\hat{U}_0$, where $\hat{\boldsymbol{U}}^2$ is given by
$\hat{\boldsymbol{U}}^2 = \frac{1}{2} \left( \hat{U}^+ \hat{U}^- + \hat{U}^- \hat{U}^+ \right) + \left( \hat{U}_0 \right)^2.$ 
\begin{eqnarray}
    &&\ket{j,m}_u = \left(-\hat{U}^-\right)^{j-m} \ket{ \mathrm{SHS}\kuo{\alpha,\gamma} }, \\
    &&\hat{\boldsymbol{U}}^2 \ket{j,m}_u = j(j+1) \ket{j,m}_u,\\
    &&\hat{U}_0 \ket{j,m}_u = m \ket{j,m}_u.
\end{eqnarray}

The Hamiltonian in \eqref{Hamhxxz} can be divided into four components, 
\begin{eqnarray}
    \hat{H}_{\rm{XXZ}} = \hat{H}_{\Delta} + h\cos{\alpha}\hat{U}_{0} - \frac{h}{2}\sin{\alpha}\left(\hat{U}^+ + \hat{U}^- \right).
\end{eqnarray}
$\left| \mathcal{S}_l \right>$ are eigenstates of $H$, so no matter how many times $H$ acts on $\left| \mathrm{SHS} \right>$, 
the product can be expanded by $\left| \mathcal{S}_l \right>$. Now we can get,
\begin{eqnarray}
    \hat{H}_\Delta \hat{H}_{\rm{XXZ}}^{n} \left| \mathrm{SHS}\kuo{\alpha,\gamma} \right> = J\Delta L \hat{H}_{\rm{XXZ}}^{n} \left| \mathrm{SHS}\kuo{\alpha,\gamma} \right>.
\end{eqnarray}

The initial state $\left| \mathrm{SHS} \right>$ have an effective spin-$L$ representation, i.e. $\left|j=L,m=L\right>_{u}$.
\begin{eqnarray}
    \hat{H}_{\rm{XXZ}} \left|L,L-n\right>_{u} =&& (J\Delta L+(L-n)h\cos{\alpha})\left|L,L-n\right>_{u} \nonumber \\ 
    &&+ \frac{h}{2}\sin{\alpha}\sqrt{n\left(2L+1-n\right)} \left|L,L-n-1\right>_{u} +   \frac{h}{2}\sin{\alpha}\sqrt{(n+1)\left(2L-n\right)} \left|L,L-n+1\right>_{u}, \qquad \qquad
\end{eqnarray}
Apply the orthogonalization process in turn to $\left\{ \left|L,L\right>_{u},\hat{H}_{\rm{XXZ}}\left|L,L\right>_{u},\cdots, \hat{H}_{\rm{XXZ}}^{2L}\left|L,L\right>_{u}\right\}$, the Krylov basis is given by
\begin{eqnarray} \label{Kbasis}
    \left\{ \left|\mathcal{K}_n\right> \right\} = \left\{ \left|L,L-n\right>_{u} ,n=0,1,2,\cdots,2L\right\}.
\end{eqnarray}
The Lanzcos coefficients is given by,
\begin{align}
    a_n &= \left< \mathcal{K}_n \right| \hat{H}_{\rm{XXZ}} \left| \mathcal{K}_n \right> = J\Delta L + h\cos{\alpha}\kuo{L-n}, \\
    b_n &= \left< \mathcal{K}_{n} \right| \hat{H}_{\rm{XXZ}} \left| \mathcal{K}_{n-1} \right> = \frac{1}{2}h\sin{\alpha}\sqrt{n\left(2L+1-n\right)}. 
\end{align}

Consider the time evolution, $\left| \phi \left(t\right) \right> =  e^{-i\hat{H}t} \left| L,L\right>$. The BCH formula gives the time evolution operator
\begin{eqnarray}
    e^{-i\hat{H}t}=e^{-iJ\Delta Lt}e^{c_+ \hat{U}^{+}}e^{c_0 \hat{U}_0}e^{c_- \hat{U}^{-}}\;
\end{eqnarray}
where

\begin{eqnarray}
c_+(t)=c_-(t)=\frac{\sin{\alpha}}{i\cot\left(ht/2\right)-\cos{\alpha}},\qquad c_0(t)=-2\ln\left[\cos{\frac{ht}{2}} + i \cos{\alpha} \sin{\frac{ht}{2}}\right].
\end{eqnarray}
The time evolution state is then
\begin{eqnarray}
    \left|{\psi(t)}\right>=&&e^{-i\hat{H}_{\rm{XXZ}}t}\ket{L,L}\nn\\
=&&e^{-iJLt}e^{-Lc_0}\sum^{2j}_{n=0}c_+^n\sqrt{\frac{(2L)!}{n!(2L-n)!}}\left|{L,L-n}\right>_u\;. \nn
\end{eqnarray}
Then it gives the time evolution state the Krylov basis coefficients in \eqref{krylovbasis},
\begin{eqnarray}
    \phi_n(t)=e^{-iJLt}e^{-Lc_0}c_+^n\sqrt{\frac{(2L)!}{n!(2L-n)!}}.
\end{eqnarray}
Starting from the $\left| \mathrm{SHS} \right>$, the fidelity exhibits periodic revivals, signaling a violation of the eigenstate thermalization hypothesis. The fidelity between the time-evolved state and the Krylov basis in Eq.(\ref{Kbasis}) is:
\begin{eqnarray}
    F_n(t) = &&\left| \left\langle \psi(t) \mid \mathcal{K}_n \right\rangle \right|^2 \nn \\
    = &&\mathbb{C}_{2L}^n\kuo{1-\sin^2{\alpha}\sin^2\frac{ht}{2}}^{\kuo{2L-n}} \kuo{\sin^2{\alpha} \sin^2\frac{ht}{2}}^{n}.
\end{eqnarray}
and the complexity is given by
\begin{eqnarray}
    C(t)=2L\sin^2{\alpha} \sin^2{\frac{ht}{2}},
\end{eqnarray}
which exhibits nonthermalization.

\section{GENERALIZATION OF SU(2) ALGEBRA AND SCAR DYNAMICS TO HIGHER DIMENSIONS} 

In this chapter, we explore how the scar phenomenon generalizes from the one-dimensional model to the two-dimensional and three-dimensional XXZ models. Through analytical derivations and numerical calculations, we demonstrate the continuation of the SU(2) algebra and scar dynamics in these higher-dimensional systems, supported by corresponding data.

\begin{figure}[H]
    \centering
    \includegraphics[width=0.5\linewidth]{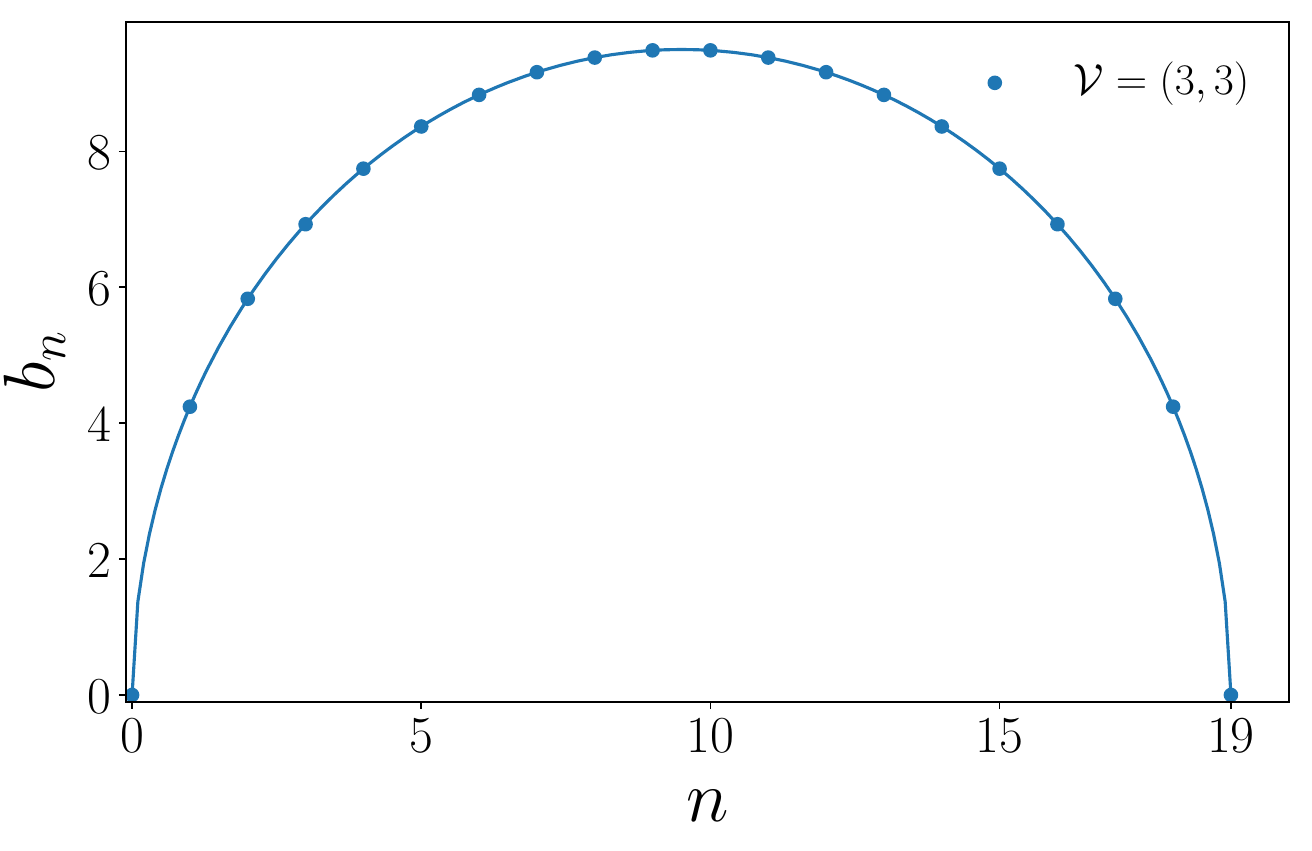}
    \caption{The Lanczos coefficients  $b_n$ versus Krylov index $n$ for two-dimensional spin helix state \eqref{generalSHS}. The system size  $\mathcal{V} = (L,L)=(3,3)$, and the parameters are $J = 1$, $h = 2$, $\alpha = \pi/2$, and $\gamma = 2\pi /L$. The dots represent the numerical results, while the solid line corresponds to the analytical expression in Eq.~\eqref{eq2bn}. The agreement between the numerical data and the analytical expression is excellent for all system sizes $\mathcal{V}$.}
    \label{Supplement fig3}
\end{figure}

The initial state we selected is the spin helix state in a two-dimensional system with size \(\mathcal{V}=(3,3)\). We find that \(b_n\) still grows elliptically with respect to \(n\), while \(a_n\) increases linearly. The value of \(b_n\) is given by:
\begin{eqnarray}  
\label{eq2bn} 
b_n = \frac{h}{2} \sin{\alpha} \sqrt{n(2L^2+1-n)} 
\end{eqnarray}
Fig. \ref{Supplement fig3} clearly shows the agreement between the numerical and analytical results for \(b_n\). When \(n = 2L^2 + 1\), the Krylov basis truncates, and the dimension of the scar subspace is \(2L^2 + 1\), which corresponds to the number of scar states. This implies that the Hamiltonian can be constructed from the generators of the SU(2) algebra.

\begin{figure}[H]
    \centering
    \includegraphics[width=0.5\linewidth]{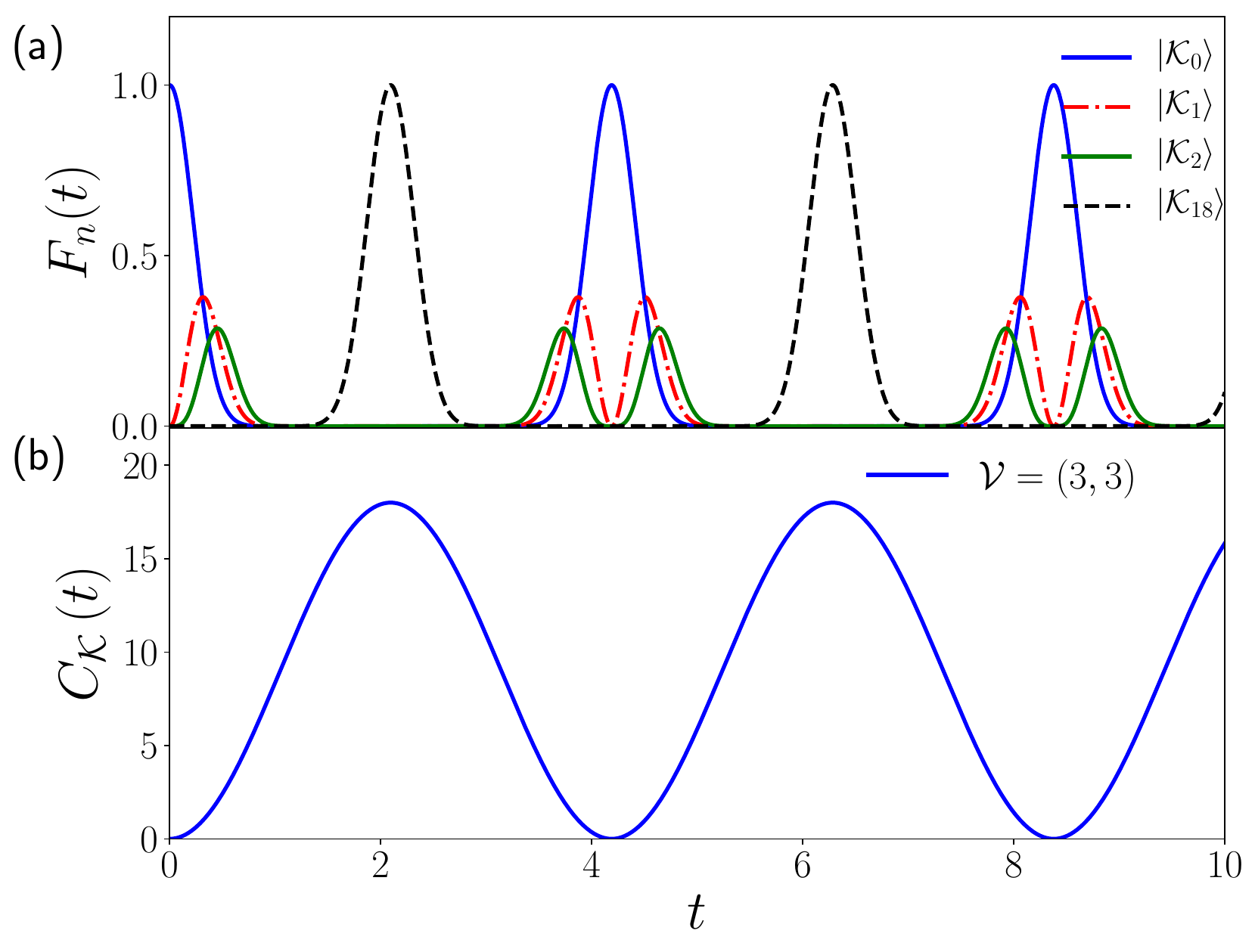}
    \caption{The dynamics of the two-dimensional XXZ model (\ref{Hamhxxz}) starting from a two-dimensional $\ket{\mathrm{SHS}\kuo{\alpha,\gamma}} $ defined in Eq.~(\ref{generalSHS}) with parameters $J = 1$, $h = 1.5$, $\alpha = \pi/2$, and $\gamma = 2\pi /L$. (a) The overlap of three Krylov vectors $\{$$ \left| \mathcal{K}_n \right>$, $n$=0, 1, 2, $2L^2\}$ with the time-evolved state.  The system size $\mathcal{V} = (L,L)=(3,3)$. 
(b) The Krylov complexity of the spin helix state.}
    \label{Supplement fig4}
\end{figure}

We continue the dynamical analysis of the 2D spin-helix state and obtain the results shown in Fig. \ref{Supplement fig4}. This exhibits the same behavior as in the main text, in Fig. \ref{FCxy} and Fig. \ref{FCxxz}. The initial state fidelity revives periodically, and the Krylov basis removes the bases with the maximum and minimum indices. Each period exhibits a double peak. The Krylov complexity now oscillates between 0 and \(2L^2\), with the expectation of the initial state diffusion depth. The analytical expression is:
\begin{eqnarray}
C(t) = 2L^2 \sin^2{\alpha} \sin^2{\frac{ht}{2}}.
\end{eqnarray}

For three-dimensional systems, numerical calculations are considerably more challenging, and therefore, we do not present explicit numerical results.  However, based on the analysis of one- and two-dimensional models, we can infer conclusions about the three-dimensional XXZ model.   In an XXZ lattice  of size $\mathcal{V} = (L,L,L)$, with the spin helix state $\ket{\rm{SHS}(\alpha,\gamma)}$ as the initial state, the following expression is obtained:
\begin{eqnarray}
&&b_n = \frac{h}{2} \sin{\alpha} \sqrt{n(2L^3 + 1 - n)}.
\end{eqnarray} 
This result extends our findings from lower-dimensional cases to the three-dimensional XXZ model, offering insight into its behavior despite the computational challenges.

\section{THE CHARACTERISTICS OF INITIAL STATES IN THE NON-SCAR SUBSPACE UNDER THE KRYLOV METHOD}
In this section, we analyze the dynamics of initial states outside the scarred subspace, focusing on their Krylov complexity and Lanczos coefficients. By comparing these results with those of scarred states, we highlight the utility of the Krylov approach in distinguishing between thermalizing and non-thermalizing behavior. 

We begin by examining the behavior of generic thermalizing initial states in the XY and XXZ models. For such states, the Krylov complexity exhibits the following characteristics: The Lanczos coefficients lack a clear analytical structure, and \(b_n\) is difficult to converge to 0. After an initial transient period, the Krylov complexity saturates to a value proportional to the system size, signaling thermalization.

\begin{figure}[H]
    \centering
    \includegraphics[width=0.8\linewidth]{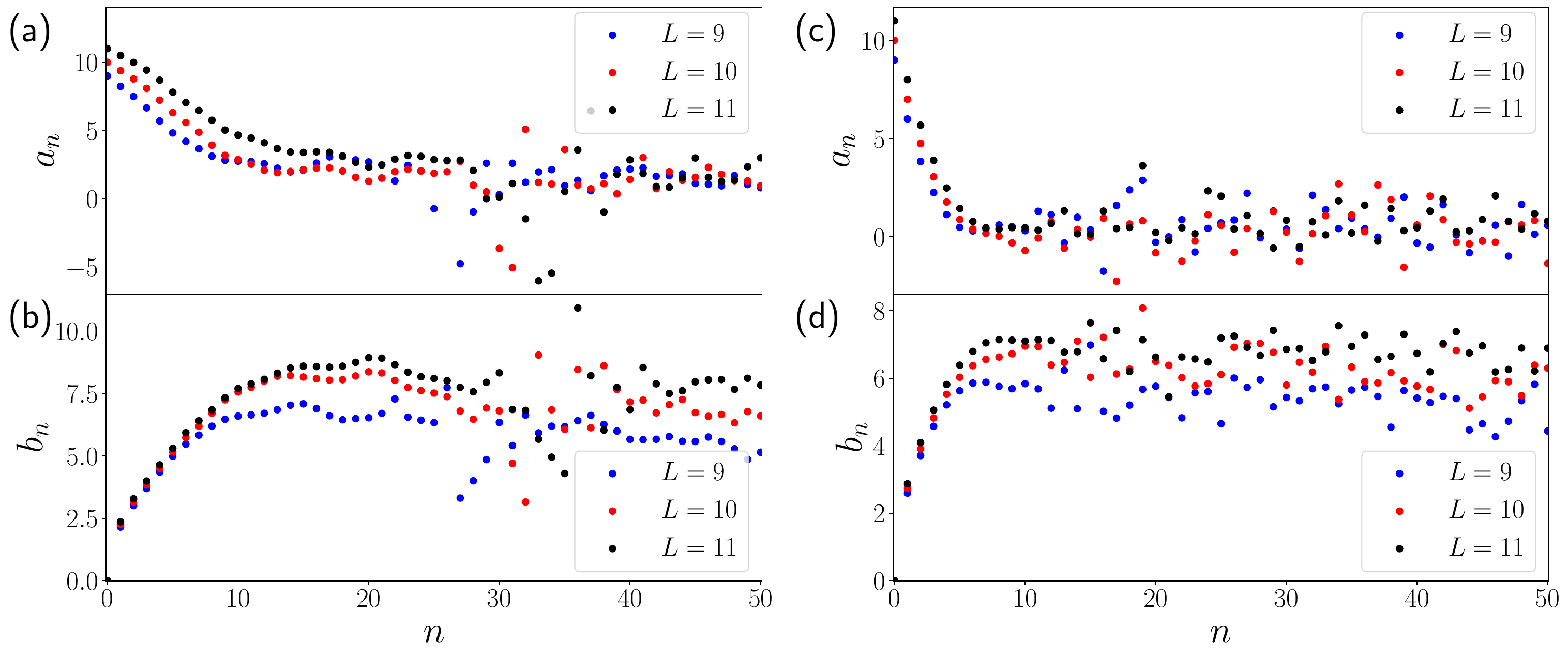}
    \caption{Lanczos coefficients for initial states \(\ket{\rm FM} \) \eqref{FM} outside the scar subspace. (a) (b) Behavior of $a_n$ and $b_n$ in the XXZ model with parameters $L=9$, $J=1$, $h=1$, and $\Delta= \cos(2\pi/9)$. (c) (d) Behavior of $a_n$ and $b_n$ in the XY model with parameters $L=9$, $J=1$, and $h=1$. }
    \label{Supplement fig1}
\end{figure}

\begin{figure}[H]
    \centering
    \includegraphics[width=0.8\linewidth]{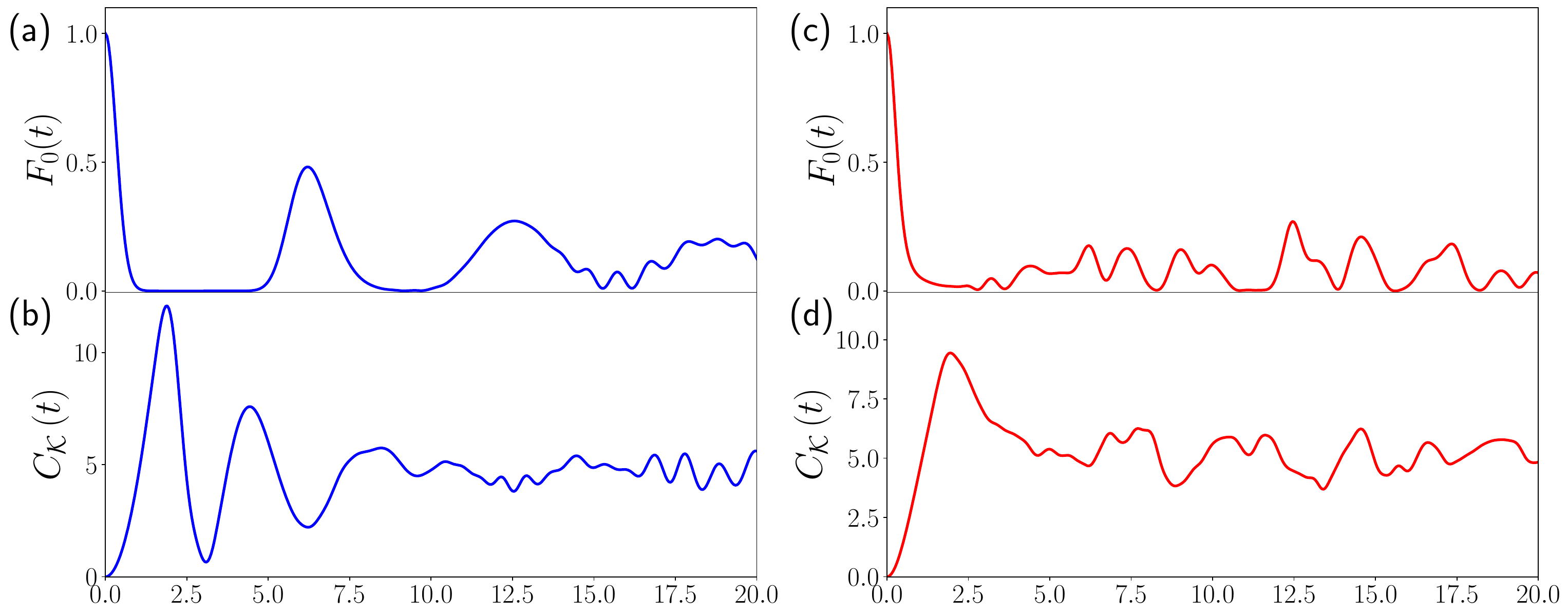}
    \caption{The Krylov results for the initial state outside the scar subspace. (a) (b) Fidelity of the initial state \( \ket{\rm FM} \) and Krylov complexity in the XXZ model, with \( J = 1 \), \( h = 1 \) and \( \Delta = \cos{2\pi}/{9} \). (c) (d) Fidelity of the initial state \( \ket{\rm FM} \) and Krylov complexity in the XY model, with \( J = 1 \) and \( h = 1 \).
     }
    \label{Supplement fig2}
\end{figure}
Outside the scar subspace, we define a simple initial state \(\ket{\rm FM}\) as follows:

\begin{eqnarray} \label{FM}
    \ket{\rm FM} = \bigotimes_{j=1}^{L} \left( \frac{\left| m_j^z = +1 \right\rangle + \left| m_j^z = -1 \right\rangle}{\sqrt{2}} \right).
\end{eqnarray}

As can be seen from Fig.~\ref{Supplement fig1}, for such a simple form of a ferromagnetic state, the Lanczos coefficients in the XXZ model are highly complex, and \(b_n\) does not easily converge. We only show the Lanczos coefficients for \(n \leq 50\). From Fig.~\ref{Supplement fig2} (a), it is evident that the fidelity of $ \ket{\rm{FM}}$ rapidly decreases and fluctuates irregularly around zero, indicating that \(\ket{\rm FM}\) thermalizes quickly in the XXZ model. Focusing on the evolution of its complexity, after a brief period of fluctuation, it relaxes to a stable value, suggesting that the initial state has diffused to a stable average depth.

\end{document}